\documentclass[a4paper,twocolumn,superscriptaddress,11pt,unpublished]{quantumarticle}
\pdfoutput=1
\usepackage[utf8]{inputenc}
\usepackage[english]{babel}
\usepackage[T1]{fontenc}
\usepackage{amsmath}
\usepackage[colorlinks,pdftex]{hyperref}

\usepackage{tikz}
\usepackage{lipsum}

\usepackage{amsthm}
\usepackage{graphicx}
\usepackage{epstopdf}
\usepackage{bm}
\usepackage{amsmath,amssymb,amsfonts}
\usepackage{xcolor}
\newtheorem{theorem}{Theorem}

\newtheorem{lemma}{Lemma}

\newtheorem{prop}{Proposition}
\usepackage{hyperref}
\usepackage{braket}

\usepackage[sort&compress,numbers]{natbib}

\usepackage[normalem]{ulem}
\newcommand\erase{\bgroup\markoverwith{\textcolor{red}{\rule[.5ex]{2pt}{0.4pt}}}\ULon}
\newcommand\eraseblue{\bgroup\markoverwith{\textcolor{blue}{\rule[.5ex]{2pt}{0.8pt}}}\ULon}

\begin{document}

\title{Effect of alternating layered ansatzes on trainability of projected quantum kernel}
\date{}
\author{Yudai Suzuki}
\affiliation{Department of Mechanical Engineering, Keio University, 3‑14‑1 Hiyoshi, Kohoku, Yokohama 223‑8522, Japan}
\author{Muyuan Li}
\affiliation{IBM Quantum, IBM-MIT Watson AI Lab, Cambridge MA 02142, USA}

\maketitle

\begin{abstract}
Quantum kernel methods have been actively examined from both theoretical and practical perspectives due to the potential of quantum advantage in machine learning tasks.
Despite a provable advantage of fine-tuned quantum kernels for specific problems, widespread practical usage of quantum kernel methods requires resolving the so-called vanishing similarity issue, where exponentially vanishing variance of the quantum kernels causes implementation infeasibility and trainability problems.
In this work, we analytically and numerically investigate the vanishing similarity issue in projected quantum kernels with alternating layered ansatzes.
We find that variance depends on circuit depth, size of local unitary blocks and initial state, indicating the issue is avoidable if shallow alternating layered ansatzes are used and initial state is not highly entangled. 
Our work provides some insights into design principles of projected quantum kernels and implies the need for caution when using highly entangled states as input to quantum kernel-based learning models.

\end{abstract}

\section{Introduction}
Recent advances in quantum devices and their public accessibility have led a number of researchers to explore the applicability of quantum computing in various fields.
Machine learning is one of such field where quantum computers can possibly enhance capability of the conventional methods. 
Remarkably, it has been shown that some quantum machine learning (QML) methods are theoretically guaranteed to outperform their existing classical counterparts for certain tasks \cite{biamonte2017quantum,rebentrost2014quantum,farhi2001quantum,liu2021rigorous,huang2021power,sweke2021quantum,huang2021information,huang2022QuantumAdvantage}.
Motivated by these works, QML approaches have also been heuristically examined with the hope to discover practical advantages over classical ones.

Quantum kernel methods are promising QML methods where the Hilbert space accessed by quantum computers are utilized as a feature space for machine learning tasks \cite{havlivcek2019supervised,schuld2019quantum}.
More specifically, quantum computers are used to map data into quantum feature space (i.e., the Hilbert space) via quantum circuits; then a quantum kernel, an inner product of a pair of data-dependent quantum features, is computed. 
The core idea is that the quantum kernel can measure the similarity between data points in the quantum feature space, without explicitly determining the corresponding feature vectors that are exponentially large in the number of qubits.
Much attention has been paid to quantum kernel methods because the provable advantage for a specific learning task has been shown \cite{liu2021rigorous} and supervised QML models can be recast in terms of kernel methods~\cite{schuld2021supervised}.

\begin{figure*}[t]
    \centering
    \includegraphics[scale=0.8]{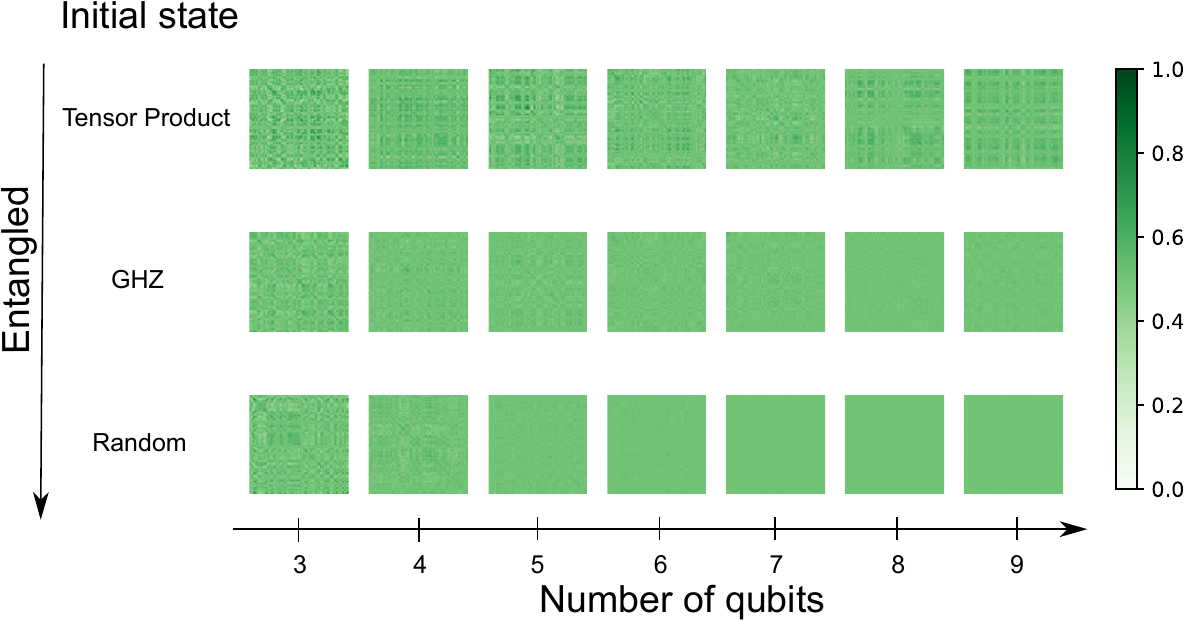}
    \caption{Gram matrices of projected quantum kernels for different number of qubits and initial states. The Gram matrices, where entry $(i,j)$ contains value of the quantum kernel for a data pair $(\bm{x}_i,\bm{x}_j)$, are computed using alternating layered ansatzes with depth $L=6$ for 100 randomly generated data points.
    Here a tensor product state (top row), the GHZ state (middle row) and a quantum state randomly sampled from the Haar measure (bottom row) are prepared as the initial state.
    The more entangled the initial state is, the more identical every element of the Gram matrix is as the the number of qubits increases; namely, vanishing similarity arises. }
    \label{fig:gram_mat}
\end{figure*}

Despite the hope of quantum advantages for real-world machine learning tasks, it has been suggested that quantum kernel methods suffer from the so-called vanishing similarity issue or exponential concentration issue \cite{suzuki2022quantum, thanasilp2022exponential}, which undermines implementation feasibility and trainability of quantum kernel-based learning models.
Analogous to the well-known barren plateau problems in variational quantum algorithms \cite{mcclean2018barren,wang2021noise,cerezo2021higher,arrasmith2021effect,cerezo2021cost}, vanishing similarity is a phenomenon where expectation value and variance of the quantum kernel decay exponentially quickly in the number of qubits.
As a result, output values of quantum kernels for any pairs of data points result in the same value, i.e., concentrated around the expectation value. 
Firstly, this implies that an exponential number of measurement shots is needed to estimate each quantum kernel on quantum hardware.
It also implies that models constructed from quantum kernels fail to distinguish the difference between data points, leading to overfitting and poor performance to new unseen data~\cite{suzuki2022quantum,thanasilp2022exponential}.

Recent works have attempted to analytically clarify the phenomenon and seek out a remedy to this issue.
In particular, Ref.~\cite{thanasilp2022exponential} analyzed the phenomenon for two types of fidelity-based quantum kernels, the commonly-used fidelity-based quantum kernel \cite{havlivcek2019supervised} and projected quantum kernels \cite{huang2021power}. 
In addition, four causes of the problem were elucidated in the literature: expressivity of quantum circuits, global measurement, how entangled the data-embedded quantum states are and quantum noise. 
The analysis gives insight into design principles for quantum kernels.
Scaling the rotation angles for data encoding gates could help avoid the issue at the cost of expressivity of quantum circuits~\cite{kubler2021inductive,canatar2022bandwidth,sim2019expressibility}.
Moreover, it has been shown that a new type of quantum kernel called the quantum Fisher kernel can mitigate the vanishing similarity issue because local similarities are measured via the information geometric quantity of quantum circuits~\cite{suzuki2022quantum}.

In this work, we further examine projected quantum kernels from the perspective of the vanishing similarity issue.
As mentioned above, Ref.~\cite{thanasilp2022exponential} analyzed projected quantum kernels for globally random quantum circuits and reached a conclusion that one cannot mitigate the exponential concentration for the quantum circuits.
On the other hand, according to Ref.~\cite{cerezo2021cost} on how to remedy the barren plateau problem, using local cost functions and the so-called alternating layered ansatzes (ALAs) possibly resolves vanishing gradients.
This suggests a possibility that projected quantum kernels can alleviate the issue because the difference of data is measured via a local quantity, i.e., reduced density matrices of the data-dependent quantum states.
Therefore, this work analytically and numerically investigates the presence of the vanishing similarity issue in projected quantum kernels for different types of quantum circuits.

To be more specific, we provide analytical expressions for expectation value and variance of projected quantum kernels using (1) $n$-qubit random quantum circuits and (2) the ALA with $m$-qubit local unitary blocks.
We assume here that globally random quantum circuits and local unitary blocks in the ALAs form $2$-designs~\cite{bremner2016average,goldberg2017complexity,harrow2023approximate,dankert2009exact,renes2004symmetric}.
With this assumption, the globally random quantum circuits fail to avoid the issue, as demonstrated in Ref.~\cite{thanasilp2022exponential}.
As for the ALAs, we find that variance of projected quantum kernels depends on not only circuit depth and size of the local unitary blocks, but also initial state.
This result indicates that variance of projected quantum kernel with shallow ALAs can avoid the vanishing similarity issue if the initial state is not highly entangled, such as a tensor product state.
Fig.~\ref{fig:gram_mat} illustrates this result.
Moreover, we observe dependence on position of the reduced density matrices (accordingly, the light-cone of the reduced subsystem) used to calculate projected quantum kernels.
This suggests that contribution of the term in the summed projected quantum kernels differs depending on position of the subsystems.  
We then validate these analytical results by performing numerical simulation.
\\

The rest of this paper is organized as follows.
We provide overview of quantum kernel methods and details of projected quantum kernels in Section \ref{sec:qkm}.
Then we elaborate the setting of our analysis in Section \ref{sec:setting}.
Our main analytical results on the vanishing similarity issue in projected quantum kernels is detailed in Section \ref{sec:main}, which is followed by numerical simulation to demonstrate examples supporting the analytical results in Section \ref{sec:numerics}.
Lastly, Section \ref{sec:discussion} discusses the implication of our results and concludes this paper.

\section{Preliminary}
\label{sec:preliminary}
In this section we first review quantum kernel methods and provide the details of projected quantum kernels.
We also introduce the settings in our analysis.

\subsection{Quantum kernel methods} \label{sec:qkm}
Quantum kernel methods measure similarity between all possible pairs of data using a function called quantum kernel.
Originally proposed was fidelity-based quantum kernel \cite{havlivcek2019supervised} defined as
\begin{equation}
\label{eq:fidelity_QK}
k_{Q}(\bm{x}_{i},\bm{x}_{j}) = \mathrm{Tr}\left[\rho\left(\bm{x}_i,\bm{\theta}\right)\rho\left(\bm{x}_j,\bm{\theta}\right)\right],
\end{equation}
where $\rho (\bm{x},\bm{\theta})= U (\bm{x},\bm{\theta})\rho_{0} U^{\dagger} (\bm{x},\bm{\theta})$ is the density matrix representation of quantum state generated by applying a unitary operator $U (\bm{x},\bm{\theta}) $ to initial state $\rho_{0}$.
The unitary operator is realized by a quantum circuit dependent on data $\bm{x}$ and tunable parameters $\bm{\theta}$, and plays a role of feature mapping; classical or quantum data are mapped to certain quantum states that have rich information on the dataset.
Note that we also introduce parameters $\bm{\theta}$, because such quantum feature map can be engineered by optimizing $\bm{\theta}$ in practical situations~\cite{lloyd2020quantum}.

Then, Gram matrix $G$ whose $(i,j)$ element corresponds to kernel function with an input pair $(\bm{x}_i, \bm{x}_j)$, i.e., $$G_{i,j}=k_{Q}(\bm{x}_{i},\bm{x}_{j}),$$ is used to perform machine learning tasks.
Typically, kernel methods are used for classification tasks in combination with support vector machines.
The classification problem is reduced to minimizing the following cost function $L(\bm{\alpha})$ with respect to the parameter $\bm{\alpha} $;
\begin{equation}
\label{eq:svm_cost}
    L(\bm{\alpha}) = - \sum_{i}^{N} \alpha_i + \frac{1}{2} \sum_{i,j}^{N} \alpha_i \alpha_j y_i y_j G_{ij}
\end{equation}
where $N$ is the number of data points and $y_i\in\{+1,-1\}$ is the label of data $\bm{x}_i$.
With optimal parameter $\bm{\alpha}^{opt}$ obtained by solving Eq.~\eqref{eq:svm_cost}, the prediction $y(\bm{x}_{new})$ of unseen data $\bm{x}_{new}$ can be written as
\begin{equation}
\label{eq:svm_pred}
    y(\bm{x}_{new})=\text{sign}\left(\sum_i \alpha_i^{opt} y_i k_{Q}(\bm{x}_{new},\bm{x}_{i})\right).
\end{equation}

While it has been proven that there exists a dataset that is not efficiently learnable by classical models but quantum kernels \cite{liu2021rigorous}, fidelity-based quantum kernels in Eq.~\eqref{eq:fidelity_QK} suffers from vanishing similarity issue: expectation and variance of the quantum kernel declines exponentially as the number of qubits increases.
More concretely, vanishing similarity issue is mathematically defined as
\begin{equation}
    {\rm Var}_{\{\bm{x},\bm{x'}\}}[k_{Q}(\bm{x},\bm{x'})]\le B, \quad B\in \mathcal{O}(c^{-n})
\end{equation}
with $c>1$ and the number of qubits $n$. 
Here, variance is taken over all possible input data pairs $\{\bm{x},\bm{x'}\}$.
We remark that, as the quantum kernel depends on the data via a quantum feature map $U(\bm{x},\bm{\theta})$, variance can be equivalently taken over $\{U(\bm{x},\bm{\theta}),U(\bm{x'},\bm{\theta})\}$ sampled from data (and parameters) dependent unitary ensemble, i.e., ${\rm Var}_{\{U(\bm{x},\bm{\theta}),U(\bm{x'},\bm{\theta})\}}[k_{Q}(\bm{x},\bm{x'})]$. 
The reason why this is detrimental is two-fold \cite{suzuki2022quantum, thanasilp2022exponential}.
One is that an exponential number of measurements must be done to precisely estimate the quantum kernel.
The other is trainability issue. 
The Gram matrix will be close to the identity matrix for a large number of qubits and thus the model of Eq.~\eqref{eq:svm_pred} obtained by minimizing the cost function in Eq.~\eqref{eq:svm_cost} would cause overfitting.

A possible remedy to this problem is projected quantum kernels proposed in Ref.~\cite{huang2021power}, where a few variations were introduced.
A simple one is linear projected quantum kernel defined as
\begin{equation} \label{eq:def_pqk_linear}
\small
    k_{PQ}^{L}(\bm{x},\bm{x'}) = \sum_{\kappa} \mathrm{Tr}\left[ \mathrm{Tr}_{\bar{S}_{\kappa}}\left[\rho(\bm{x},\bm{\theta}) \right] \mathrm{Tr}_{\bar{S}_{\kappa}}\left[\rho(\bm{x'},\bm{\theta}) \right]\right]
\end{equation}
where $S_{\kappa}$ denotes subspace for the $\kappa$-th qubit and $\mathrm{Tr}_{\bar{S}_{\kappa}}\left[ \cdot \right]$ is partial trace operation over the subspace $\bar{S}_{\kappa}$.
Note that $\bar{S}$ is the compliment of the subspace $S$.
Also, the Gaussian projected quantum kernel is proposed:
\begin{equation} \label{eq:def_pqk_gauss}
\small
\begin{split}
    & k_{PQ}^{G}(\bm{x},\bm{x'})\\ & = \exp \left( -\gamma \sum_{\kappa} \|\mathrm{Tr}_{\bar{S}_{\kappa}}\left[\rho(\bm{x},\bm{\theta}) \right]- \mathrm{Tr}_{\bar{S}_{\kappa}}\left[\rho(\bm{x'},\bm{\theta}) \right]\|_2^2 \right)  \\
\end{split}
\end{equation}
with a hyperparameter $\gamma\in\mathbb{R}^{+}$ and the Hilbert–Schmidt norm $\|\cdot\|_{2}$.
A key point of projected quantum kernels is that similarity of data is measured using reduced density matrix $\mathrm{Tr}_{\bar{S}_{\kappa}} [\rho(\bm{x},\bm{\theta})]$ instead of the density matrix $\rho(\bm{x},\bm{\theta})$.
Namely, local difference between data is compared in projected quantum kernels. 
According to Ref.~\cite{cerezo2021cost}, the barren plateau problem in variational quantum algorithms can be circumvented using local cost functions and the ALA.
Similarly, projected quantum kernels also possess a local property that can help mitigate the vanishing similarity issue, which makes it favorable over traditional quantum kernels for practical applications.

\subsection{Setting in our analysis} \label{sec:setting}

\begin{figure*}[t]
    \centering
    \includegraphics[scale=0.8]{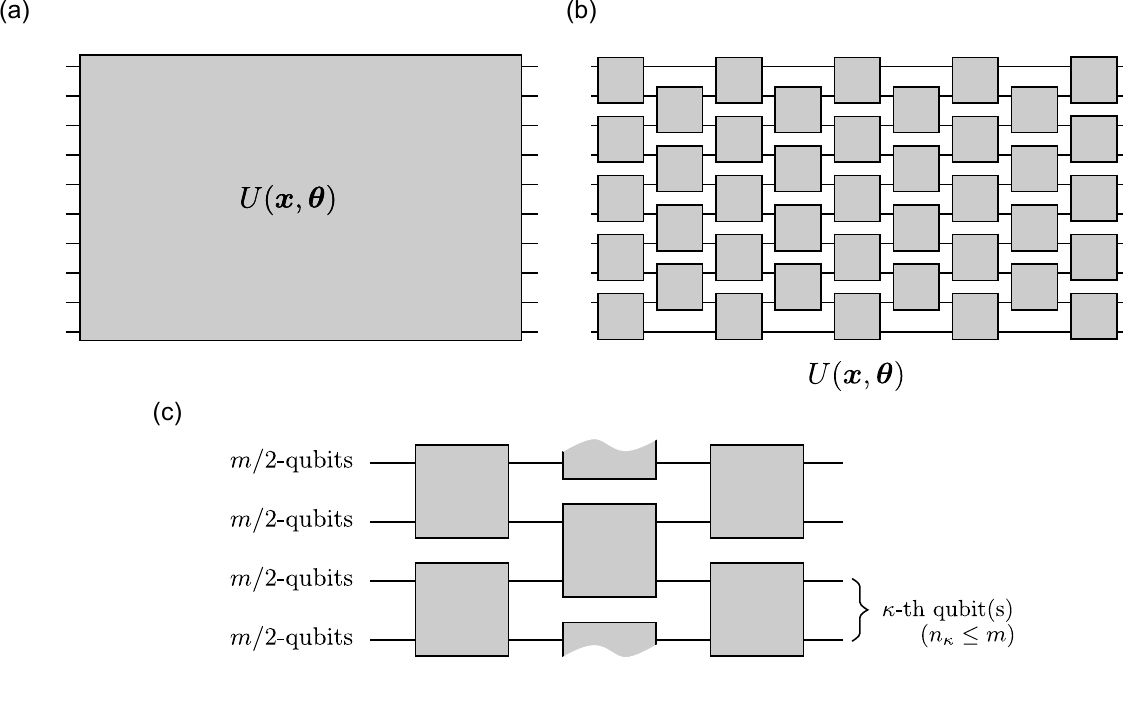}
    \caption{Quantum circuits used in our analysis. Panels (a) and (b) show the globally random quantum circuit acting on all qubits and ALA, respectively. Panels (c) shows details of ALA and the setting of projected quantum kernel in our analysis.  }
    \label{fig:q_circuit}
\end{figure*}

Although Ref.~\cite{thanasilp2022exponential} demonstrates that projected quantum kernels with globally random quantum circuits cannot avoid the issue, it is a seemingly promising approach because of their locality.
Thus, this work further analyzes projected quantum kernels from the vanishing similarity perspective, considering two types of quantum circuits.
One is the $n$-qubit random quantum circuit and the other is the ALA with $m$-qubit local unitary blocks \cite{cerezo2021cost}, as depicted in Fig.~\ref{fig:q_circuit} (a) and (b), respectively.
Let us note that the former quantum circuit is the same setting in Ref.~\cite{thanasilp2022exponential}, but the latter has not been examined for use of projected quantum kernels.
We performed analytical calculation for the globally random quantum circuits as well to make sure of the validity of our analysis and show an exact expression of the variance.
For ease of analytical investigation, we then assume that the globally random quantum circuits and local unitary blocks in the ALAs are independent and $2$-designs \cite{bremner2016average,goldberg2017complexity,harrow2023approximate,dankert2009exact,renes2004symmetric}, meaning that the quantum circuits (unitary blocks) have the same statistical property with Haar random unitary up to the second moment. 
In a broad sense, this assumption indicates that the quantum circuits or unitary blocks are expressive enough to uniformly explore the ensemble of Haar random states.
We remark that, while quantum circuits might not be $2$-designs in practice, some previous works have made similar assumptions to check the problems such as barren plateau \cite{mcclean2018barren,cerezo2021cost,holmes2021barren,holmes2022connecting,arrasmith2022equivalence} and vanishing similarity \cite{suzuki2022quantum, thanasilp2022exponential,kubler2021inductive}.  
Specifically, we express the ALA as 
\begin{equation}
\label{eq:ALA_unitary}
\begin{split}
    U(\bm{x},\bm{\theta}) &= \prod_{d=1}^{L} V_{d}(\bm{x},\bm{\theta}) \\
    &= \prod_{d=1}^{L} \left(\prod_{l=1}^{\zeta} W_{l,d}(\bm{x},\bm{\theta}_{l,d})\right),
\end{split}
\end{equation}
where $L$ is circuit depth and $\zeta$ is the number of unitary blocks in each layer.
Here we assume that the total number of qubits $n$ satisfies $n=m\zeta$.
We note that the number of qubits on which both a unitary block in a layer and the one in the adjacent layer act is $m/2$; for example, $S_{(2,1)}$ and $S_{(1,2)}$ have $m/2$-qubit subspace in common, where $S_{(l,d)}$ is the subspace of qubits which the unitary block $W_{l,d}$ acts on.
The detail is illustrated in Fig.~\ref{fig:q_circuit} (c).

Throughout this manuscript, in lieu of Eqs.~\eqref{eq:def_pqk_linear} and \eqref{eq:def_pqk_gauss}, we consider the following quantity;
\begin{equation} \label{eq:def_pqk}
    k_{PQ}^{(\kappa)}(\bm{x},\bm{x'}) = \mathrm{Tr}\left[ \mathrm{Tr}_{\bar{S}_{\kappa}}\left[\rho(\bm{x},\bm{\theta}) \right] \mathrm{Tr}_{\bar{S}_{\kappa}}\left[\rho(\bm{x'},\bm{\theta}) \right]\right].
\end{equation}
We focus on this quantity because exploring it is sufficient to confirm the tendency of projected quantum kernels.
Of course, the variance of Eq.~\eqref{eq:def_pqk_linear} depends on the covariance terms and thus is not necessarily equal to that of the summation of Eq.~\eqref{eq:def_pqk} over possible $\kappa$.
However, in this case, every covariance term is equal to or more than zero and the difference between them does not matter in terms of scaling; see Appendix \ref{ap: variance_sum} for more details.
Moreover, without loss of generality, we assume that the subspace for the $\kappa$-th reduced density matrix (composed of $n_{\kappa}$ qubits) appearing in Eq.~\eqref{eq:def_pqk}, $S_{\kappa}$, is completely included in the subspace on which one of the unitary blocks in the last layer acts, as is shown in Fig.~\ref{fig:q_circuit} (c). 
We also assume that initial state $\rho_{0}$ is an arbitrary pure state.

\section{Results}
\label{sec:result}
In what follows, we provide analytical results on the vanishing similarity issue in projected quantum kernels. 
Then, we show numerical results to check the reliability of our analysis.

\subsection{Main results} \label{sec:main}
We analytically calculate expectation value and variance of projected quantum kernels to check the existence of vanishing similarity issue.
Here, we focus on two types of quantum circuits, that is, globally random quantum circuits and ALAs.
Although the case for globally random quantum circuits has been analyzed in a previous study \cite{thanasilp2022exponential}, we here check to confirm our analytical procedure and give an exact expression of the variance.

We first show analytical results for the globally random quantum circuits with the full proof included in Appendix \ref{ap: case_random}.
\begin{prop} \label{prop:prop1}
Let us denote expectation value and variance of projected quantum kernel defined in Eq.~\eqref{eq:def_pqk} with $n$-qubit random quantum circuits as $\braket{k_{PQ,rqc}^{(\kappa)}}$ and ${\rm{Var}}[k_{PQ,rqc}^{(\kappa)}]$, respectively.
If the $n$-qubit random quantum circuits form $t$-designs with $t\ge 2$ and independent, then we have
\begin{equation}\label{eq:exp_rqc}
    \braket{k_{PQ,rqc}^{(\kappa)}}=\frac{1}{2^{n_{\kappa}}},
\end{equation}
\begin{equation}\label{eq:var_rqc}
    {\rm{Var}}[k_{PQ,rqc}^{(\kappa)}]=\frac{2^{2n_{\kappa}}-1}{2^{2n_{\kappa}}\left(2^{n}+1\right)^2}\approx \frac{1}{2^{2n}}.
\end{equation}
\end{prop}
We remind the readers that $n_{\kappa}$ is the number of $\kappa$-th qubit(s) and $n$ is the total number of qubits.
Proposition \ref{prop:prop1} implies that the similarity between a pair of different data will be hard to distinguish regardless of the size of reduced density matrix for a large number of qubits.
Therefore, projected quantum kernels with globally random quantum circuits cannot avoid vanishing similarity issue.
Note that the result is different from the previous result in Ref.~\cite{thanasilp2022exponential} in a sense that we calculate the exact expectation rather than its upper bound, but the implication is consistent.

Next, we provide the result obtained for the case of ALAs.
We here obtain the lower bound of variance to see the absence of vanishing similarity issue. 
Please refer to Appendix \ref{ap: case_ala} for the proof.
\begin{theorem} \label{thm:pqk}
For projected quantum kernel defined in Eq.~\eqref{eq:def_pqk} and ALA defined in Eq.~\eqref{eq:ALA_unitary}, we denote its expectation value and variance as $\braket{k_{PQ,ala}^{(\kappa)}}$ and ${\rm{Var}}[k_{PQ,ala}^{(\kappa)}]$, respectively.
Also, we assume that every unitary block in the ALAs,
$U(\bm{x},\bm{\theta})$ and $U(\bm{x'},\bm{\theta})$, is a $t$-design with $t\ge 2$ and independent.
Then the expectation value is
\begin{equation}\label{eq:exp_ala}
    \braket{k_{PQ,ala}^{(\kappa)}}=\frac{1}{2^{n_{\kappa}}}.
\end{equation}
As for the variance, its lower bound is
\begin{equation}\label{eq:var_ala_lower}
\small
    {\rm{Var}}[k_{PQ,ala}^{(\kappa)}] \ge
    \frac{2^{2m(L-1)}\left(2^{2n_{\kappa}}-1\right)}{(2^{2m}-1)^{2}(2^m+1)^{4(L-1)} 2^{2n_{\kappa}}} F\left(\rho_{0}, L\right),
\end{equation}
with a function $F(\rho_{0}, L)$ of the initial state $\rho_{0}$ and the depth $L$.
More specifically, we define the function as
\begin{equation} \label{eq:f_function}
\begin{split}
&F\left(\rho_{0},L \right)\\ 
&= \left(2^{m} \sum_{h\in P(S_{(k_u,1)}:S_{(k_l,1)})} t_{h} \mathrm{Tr}\left[\rho_{0,\bar{h}}^2\right] -\sum_{\tau=0}^{L-1} \frac{c_\tau}{2^{m\tau}}, \right)^2, \\
\end{split}
\end{equation} 
where $t_{h},c_{\tau}\in \mathbb{R}^{+}$ and $\rho_{0,\bar{h}} = \mathrm{Tr}_{\bar{h}}\left[\rho_0\right]$ is the partial trace of the initial state over the subspace $\bar{h}$.
Also, $P(S_{(k_u,1)}:S_{(k_l,1)})$ is the set containing all the possible neighboring subspaces in $\bigcup_{i=0}^{k_l-k_u} S_{(k_u+i,1)}$.
Here, $W_{k_{u},1}$ ($W_{k_{l},1}$) denotes the unitary block located at the upper (lower) edge of the light-cone in the first layer.
We note that $F(\rho_0,L)=0$ if $\rho_{0,\bar{h}}$ is the completely mixed state for all subspaces $h$.

\end{theorem}
Like the case for globally random quantum circuit, the expectation value is not dependent on the total number of qubits but the system size of the reduced density matrix.
However, Eq~\eqref{eq:var_ala_lower} shows that the lower bound of the variance depends not only on the depth $L$ and the size of the local unitary blocks $m$, but also the initial state via the function $F(\rho_{0},L)$. 
As shown in Eq.~\eqref{eq:f_function}, the function contains purity of some subspace of initial state.
Thus, depending on the choice of initial state, vanishing similarity issue can be avoided.
For example, if initial state can be represented as a tensor product of arbitrary single-qubit state, i.e., $\rho_0=\sigma_{1}\otimes\sigma_{2}\otimes\ldots\otimes\sigma_{n}$ with arbitrary single-qubit states $\{\sigma_i\}$, then the function has a maximum value and the variance scales $\Omega(2^{-2mL})$.
On the other hand, if initial state is so entangled that $\mathrm{Tr}[\rho_{0,\bar{h}}^2]$ is the completely mixed state for almost all $h$, then the variance could decrease exponentially fast with respect to the number of qubits regardless of circuit depth.
Note that it has been reported that initial state matters for the vanishing gradient problems in variational quantum algorithms \cite{leone2022practical,larocca2022diagnosing}.
Thus, our result suggests that initial state should also be taken into account for the usage of projected quantum kernels.

\begin{figure}[t]
    \centering
    \includegraphics[scale=0.6]{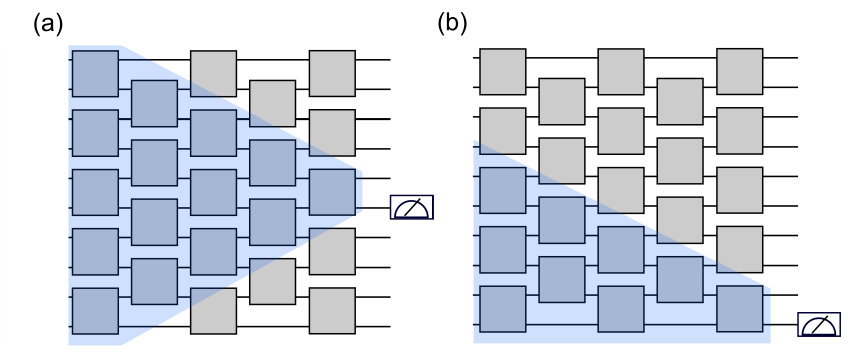}
    \caption{Light-cone depending on the position of $\kappa$-th qubit(s). The blue regions represent the light-cone in the quantum circuits. Panel (a) shows the case ($\mathrm{i}$) where the number of local unitary blocks in the light-cone is the largest, while Panel (b) shows the case ($\mathrm{ii}$) with the smallest number of unitary blocks. }
    \label{fig:light cone}
\end{figure}

\begin{figure*}[t]
    \centering
    \includegraphics[scale=0.8]{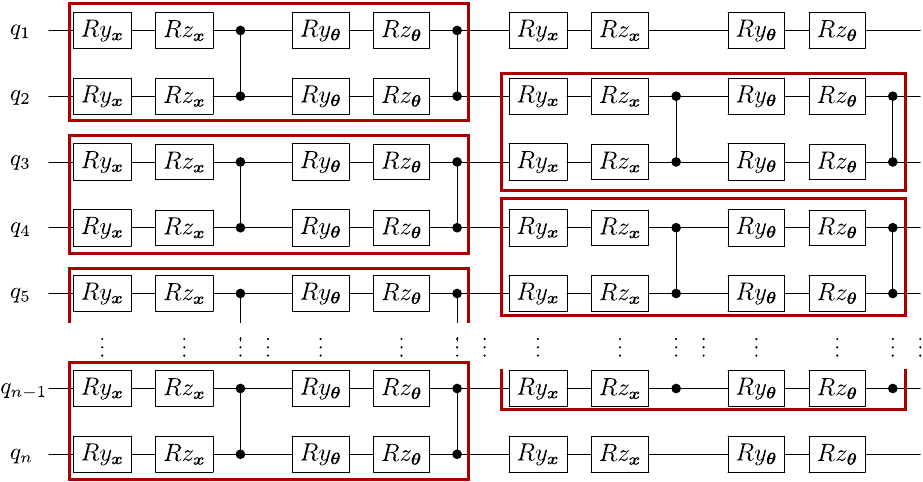}
    \caption{Alternating layered ansatz used in our simulation. As an example, here we show an even $n$-qubit alternating layered ansatz with depth $L=2$.
    The quantum circuit consists of $2$-qubit local unitary blocks denoted by red boxes, each of which has a data embedding layer and a parameterized quantum circuit layer.
    We note that $Ry_{a}$ ($Rz_{a}$) represents a single-qubit rotation gate on $Y$ ($Z$) axis, whose angle is determined by a function of $\bm{x}$ or $\bm{\theta}$ shown in the subscript.
    In the numerical experiments, $i$-th element of data $x_{i}$ is encoded into single-qubit rotation gates ($Ry$ and $Rz$) acting on the $i$-th qubit in every embedding layer. Also, each parameter is assigned to different single-qubit rotation gates in the parameterized quantum circuits layers.}
    \label{fig:q_circuit_numerics}
\end{figure*}

Moreover, we check dependence of the variance on position of the $\kappa$-th qubit.
To be more specific, we consider the following situations; the $\kappa$-th qubit(s) is (are) located ($\mathrm{i}$) in the middle of the last layer and ($\mathrm{ii}$) in the unitary block at the edge i.e., $W_{1,L}$ or $W_{\zeta,L}$.
Fig.~\ref{fig:light cone} (a) and (b) illustrate these cases respectively. 
In addition, we assume that initial state is a tensor product state to check the relationship between the depth and the position of $\kappa$-th qubit(s).
In the first case, this is exactly the same as the result shown in Eq~\eqref{eq:var_ala_lower}, i.e., $\Omega(2^{-2mL})$.
For the second case, as demonstrated in Appendix \ref{ap: case_ala}, the variance is $\Omega(2^{-mL})$.
The difference comes down to the number of unitary blocks in the light-cone.
This implies that reduced density matrices at the edge of the layer contribute to the linear projected quantum kernel in Eq.~\eqref{eq:def_pqk_linear} more than the ones in the middle due to the quadratic difference. 
We remark that the dependence of variance on the observables' position was argued in the context of variational quantum algorithms in Ref.~\cite{cerezo2021cost}, and the result we newly obtained here from the viewpoint of quantum kernel methods is similar to the statement shown in the literature; see Supplementary Information Figure. 2 of Ref.~\cite{cerezo2021cost}.

\subsection{Numerical results}
\label{sec:numerics}

We perform numerical simulations to demonstrate examples that support our analytical results.
In particular, we focus on the behavior of variance for the ALA, because the one for the globally random quantum circuits has been analyzed in Ref.~\cite{thanasilp2022exponential}.
In the numerical experiments, ALAs with $2$-qubit local unitary blocks shown in Fig.~\ref{fig:q_circuit_numerics} are considered, where we employ data re-uploading techniques \cite{perez2020data}.
Namely, each local unitary block consists of embedding layer and the parameterized quantum circuit layers.
Here, we use rotation $Y$ and $Z$ gates as single-qubit rotation gates acting on the $i$-th qubit, i.e.,  $R_{\sigma_{i}}(\beta)=\exp(-\beta \sigma_i/2), \sigma_{i}\in \{Y_i,Z_i\}$, and the controlled-Z gate as an entangler.
As for the input data, we set the number of qubits equal to the dimension of the data and each component is randomly chosen from the uniform distribution ranging $[-\pi,\pi)$.
Analogously, each parameter in the parameterized quantum circuit layers are selected uniformly at random from the range $[-\pi,\pi)$.
Then we prepare five sets of parameters and five datasets containing $50$ data paints to compute $k_{PQ}^{(\kappa)} (\bm{x},\bm{x'})$ in Eq.~\eqref{eq:def_pqk} with $\bm{x} \neq \bm{x'}$.
We note that $n_{\kappa}=1$ for our numerical simulations.
Afterwards, the variance is calculated using the projected quantum kernels computed for different 25 settings of input dataset and the parameter set.
The computation is performed for all possible $\kappa$.
When we encode the data into the quatum circuit, the $i$-th component of the input data, $x_i$, is injected into the angle of the single-qubit rotation gates acting on the $i$-th qubit in every embedding layer; that is, $R_{Y_{i}}(x_i)$ ($R_{Z_{i}}(x_i)$). 
We also assign each parameter to a single-qubit rotation gate in parameterized quantum circuit layers.
Namely, no parameters are shared with different rotation gates. 
Fig.~\ref{fig:q_circuit_numerics} depicts the details of the quantum circuit. 
The numerical simulation is performed using Qiskit~\cite{Qiskit}.\\

We here summarize the numerical results from the following perspectives: ($\mathrm{i}$) the dependence of variance on circuit depth for different initial states, ($\mathrm{ii}$) the dependence on position of the $\kappa$-th qubit and ($\mathrm{iii}$) the relation between the variance and the number of qubits $n$.

\begin{figure}[t]
    \centering
    \includegraphics[scale=0.45]{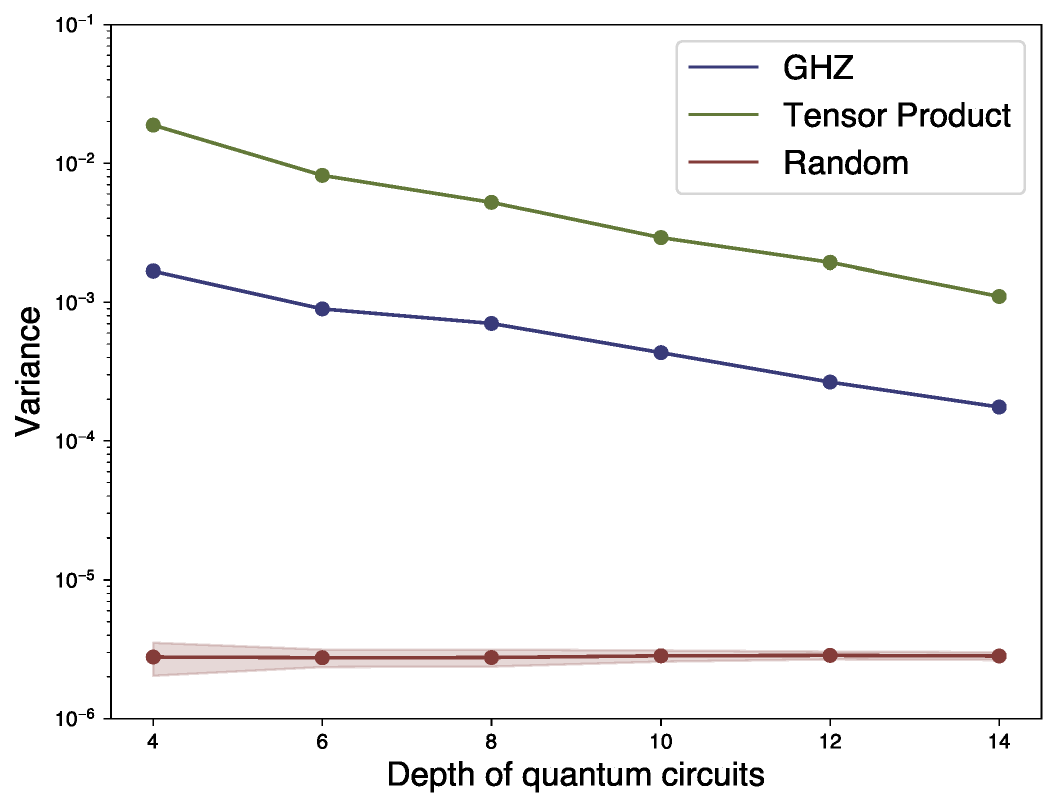}
    \caption{Variance of projected quantum kernel against the depth of quantum circuits. Here we used the $9$-qubits ALAs with the depth $L\in \{4, 6, 8, 10, 12, 14\}$, and the reduced density matrix for the fifth qubit to compute the projected quantum kernel. 
    We consider three initial states: a tensor product state (green), the GHZ state (blue) and random quantum states (red). The shaded region illustrates the standard deviation over five different random states.}
    \label{fig:var_dep}
\end{figure}

\subsubsection{Dependence on circuit depth}
Fig.~\ref{fig:var_dep} shows the variance of projected quantum kernels against the depth $L$ for different initial states, where the number of qubits $n=9$ and the reduced density matrix with respect to the fifth qubit is considered for three initial states: a tensor product state $\rho_{0}=\ket{0^{\otimes n}}\bra{0^{\otimes n}}$, the GHZ state $\rho_0=\ket{\psi_{GHZ}}\bra{\psi_{GHZ}}$ with $\ket{\psi_{GHZ}}=2^{-1/2}(\ket{0}^{\otimes n}+\ket{1}^{\otimes n})$ and initial states randomly sampled from the Haar measure.
We choose these initial states with different degrees of entanglement to examine how entanglement of initial states affects the variance.
As for the random initial states, we prepare five different states and the variance is averaged over the trials.
It turns out that the variance decreases exponentially in circuit depth $L$ for the case of the tensor product state and the GHZ state. 
On the other hand, if a random quantum state is prepared as the initial state, the variance is independent of the depth and much smaller than the ones for other cases.
This is consistent with the analytical result shown in Theorem \ref{thm:pqk}.
As demonstrated in Eq.~\eqref{eq:var_ala_lower}, the variance is determined by the depth and the function of the initial state $F(\rho,L)$.
For the first two cases, $F(\rho,L)$ does not contribute to the variance so much because the reduced systems of the initial states are far from the completely mixed states; the purity is one for the tensor product state over any subspace $h$, and the purity is $1/2$ for the GHZ state if $\bar{h} \neq \emptyset$ or $h\neq\emptyset$ and otherwise one.
Thus, the term other than $F(\rho,L)$ comes into play; the variance vanishes exponentially with respect to the depth.
Yet, the partial trace of a random quantum state can be close to the completely mixed state and thus $F(\rho,L)$ plays an significant role in the variance rather than the remaining term.
Hence, the variance is consistently small regardless of the depth.

\begin{figure*}[t]
    \centering
    \includegraphics[scale=0.8]{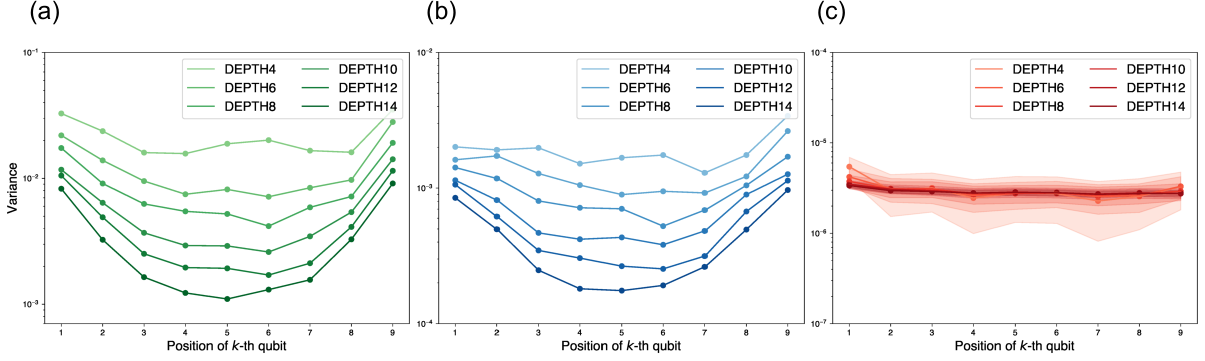}
    \caption{Variance of projected quantum kernel against the position of $\kappa$-th qubit. We used the ALAs with $9$-qubits. The variances of projected quantum kernel with depth $L\in \{4, 6, 8, 10, 12, 14\}$ are shown for the case of (a) a tensor product state, (b) the GHZ state and (c) random quantum states.
    In panel (c), the standard deviation is represented by the shaded region.}
    \label{fig:var_pos}
\end{figure*}

\subsubsection{Dependence on positions of reduced subsystems}
The variance against positions of the $\kappa$-th qubit for $9$ qubits system is shown in Fig.~\ref{fig:var_pos}.
We notice that the variance of the reduced system at the edge of the layer is smaller than that of the systems in the middle for the tensor product state and the GHZ state, shown in Fig.~\ref{fig:var_pos} (a) and (b), respectively. 
Also, the gap of the variance between the systems at the edge and in the middle gets larger as the depth increases.
This numerical result agrees with the statement in the previous section that the scaling of the variance differs depending on the number of local unitary blocks in the light-cone, and accordingly the position of the $\kappa$-th qubit.
As for the random quantum state case in Fig.~\ref{fig:var_pos} (c), the depth and the position are less significant in the variance because the term $F(\rho,L)$ contributes dominantly.

\begin{figure*}[t]
    \centering
    \includegraphics[scale=0.8]{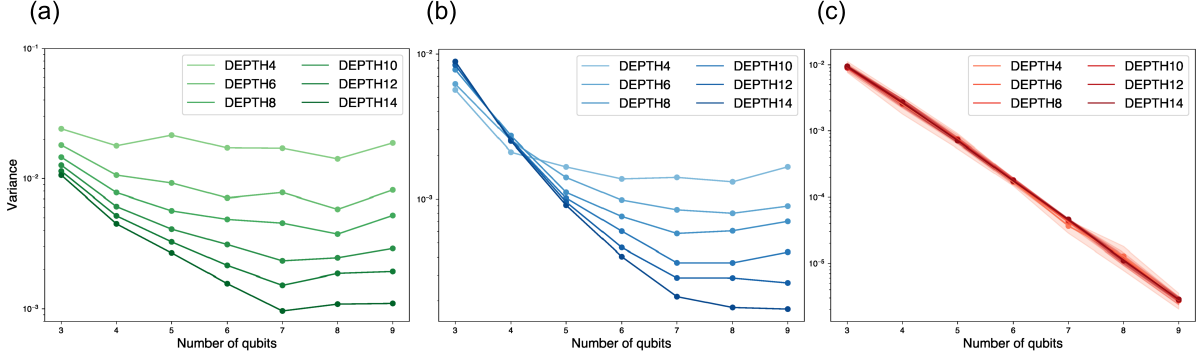}
    \caption{Variance of projected quantum kernel against the number of qubits. We used the different number of qubits, $n\in\{3, 4, 5, 6, 7, 8, 9\}$ and the ALAs with the different depth $L\in \{4, 6, 8, 10, 12, 14\}$.
    Here, we consider the reduced system of $\lceil n/2 \rceil$-th qubit, i.e., the qubit in the middle of the width.
    Panels (a), (b) and (c) show the variances for a tensor product state, the GHZ state and random quantum states, respectively.}
    \label{fig:var_qubit}
\end{figure*}

\subsubsection{Dependence on the total number of qubits}
Fig.~\ref{fig:var_qubit} (a) to (c) show the variance for the different number of qubits using a tensor product state, the GHZ state and random quantum states, respectively.
For the tensor product and the GHZ state, the variance levels off for all cases of circuit depth when the number of qubits is larger than certain number.
This is because the purity is constant for these cases and thus $F(\rho,L)$ is saturated.
Thus, we can confirm that the variance of these cases is irrelevant to the number of qubits.
However, Fig.~\ref{fig:var_qubit} (c) shows that the variance vanishes exponentially fast in the number of qubits.
This would be attributed to the fact that there is an exponential decay in $F(\rho,L)$ with respect to the number of qubits.
Hence, this indicates that initial states really matter in the variance of projected quantum kernels.

\section{Discussion \& Conclusion}\label{sec:discussion}
In this paper we investigated vanishing similarity issue in projected quantum kernels from both analytical and numerical perspectives.
We analytically showed that this issue is not avoidable for the case of globally random quantum circuits, which is consistent with previous results in Ref.~\cite{thanasilp2022exponential}.
In contrast, we found that projected quantum kernels with ALAs can avoid exponential decay of variance if the quantum circuits are shallow and initial state is not highly entangled.
This implies the potential of projected quantum kernels for practical usability.
In addition, we showed that initial state plays a significant role in the variance scaling and thus caution needs to be taken for preparing input states.
We discuss the implication of our results in QML tasks below.

First, our results suggest that there is a caveat when quantum data is used as input states in QML tasks.
Some QML tasks handle quantum states as an input state, and then parameterized quantum circuits are applied to the state to seek out features suitable for the tasks.
In this situation, the initial state could be more entangled than a tensor product state.
Hence, there is a possibility that the vanishing similarity issue for projected quantum kernels could be exacerbated for some tasks.

We also showed that the variance differs depending on the position of reduced density matrix.
Thus, the contribution to projected quantum kernels in Eqs.~\eqref{eq:def_pqk_linear} and ~\eqref{eq:def_pqk_gauss} of reduced systems at the edge of the layer is larger than that of systems in the middle; the tendency gets worse as we increase circuit depth.
This might result in poor performance to some tasks because the relevant information could be undermined.
Hence, in some situations, it would be better to consider the gap, for example, by modifying weights of projected quantum kernel for the $\kappa$-th qubit(s).

Moreover, our results indicate a situation where classical shadow can reduce quantum resources required to compute projected quantum kernels. 
Classical shadow is a technique to estimate properties of quantum states with a small number of measurement shots~\cite{huang2020predicting}. 
Thus far, some works have also used the technique for quantum kernel methods~\cite{huang2021power,huang2022provably}. 
On the other hand, classical shadow does not work when vanishing similarity arises. 
This is because the resolution needed to tell the difference in a pair of data through the quantum kernel is significantly high. 
Our Theorem~\ref{thm:pqk} suggests that projected quantum kernels can utilize the power of classical shadows, when shallow ALAs are used and initial state is not highly entangled.

We lastly remark that our analytical results are based on the assumption that quantum circuits and the local unitary blocks in the ALAs are $2$-designs.
This result is of significance in that we shed light on trainability and limitations of projected quantum kernels in general.
On the other hand, as the no free lunch theorems \cite{adam2019no,wolpert1997no,wolpert2002supervised} suggest, domain knowledge should be incorporated into the model.
Actually, an emerging field called geometric quantum machine learning \cite{larocca2022group,ragone2022representation,schatzki2022theoretical,nguyen2022theory,meyer2023exploiting}, where inductive bias such as symmetry is considered in constructing quantum models, has attracted much attention.
Therefore it would be worthwhile to explore the existence of vanishing similarity issue by incorporating domain knowledge into the model for practical purpose.
It would also be important to investigate advantages of projected quantum kernels for practical machine learning tasks handling quantum data as well as classical data. \\

\textbf{Acknowledgement}
The authors thank Kunal Sharma, Ryan Sweke, Khadijeh Najafi and Antonio Mezzacapo for stimulating discussions and comments on the manuscript. Part of this work was done when Y.S. was a research intern at IBM.
Y.S. was supported by Grant-in-Aid for JSPS Fellows 22KJ2709.

\bibliographystyle{unsrtnat}
\bibliography{bibref}

\onecolumn\newpage
\appendix

This appendix is organized as follows. 
In Appendix \ref{ap: preliminary}, we show preliminary on the integration over the Haar random unitary used in our analysis.
Then, we provide the proof of our analytical results on expectation value and variance of projected quantum kernels for the case of (1) $n$-qubit random quantum circuits and (2) alternating layered ansatzes with $m$-qubit local unitary blocks, in Appendix \ref{ap: case_random} and \ref{ap: case_ala}, respectively.
In addition, we explain the difference between Eq.~\eqref{eq:def_pqk_linear} and Eq.~\eqref{eq:def_pqk} in the variance in Appendix \ref{ap: variance_sum}.

\section{Preliminaries}
\label{ap: preliminary}
We utilize formulae of integration over Haar random unitary to calculate expectation value and variance of projected quantum kernels (PQKs).
Hence we here present Lemmas related to the calculation.

\subsection{Formulas of integrals over Haar random unitaries}
\label{ap: haar random}
Our analysis assumes that quantum circuits form $t$-designs \cite{bremner2016average,goldberg2017complexity,harrow2023approximate,dankert2009exact,renes2004symmetric}.
When a quantum circuit $W$ is a $1$-design, i.e., the ensemble have the same statistical properties with the Haar random unitary up to the first moment, we can have the following expression;
\begin{equation}
\label{eq:1d_int}
    \int d\mu(W) W_{i,j}W^{*}_{l,k} = \frac{\delta_{i,l}\delta_{j,k}}{d},
\end{equation}
where $d$ is the dimension of the unitary $W$ and $\delta_{i,j}$ represents the Kronecker delta.
Similarly, we can exploit the following formula for the $2$-design case;
\begin{equation}
\label{eq:2d_int}
\begin{split}
    \int d\mu(W) W_{i_1,j_1}W^{*}_{l_1,k_1}W_{i_2,j_2}W^{*}_{l_2,k_2} &=\frac{\delta_{i_1,l_1}\delta_{i_2,l_2}\delta_{j_1,k_1}\delta_{j_2,k_2}+\delta_{i_1,l_2}\delta_{i_2,l_1}\delta_{j_1,k_2}\delta_{j_2,k_1}}{d^2-1}\\
    & \qquad -\frac{\delta_{i_1,l_1}\delta_{i_2,l_2}\delta_{j_1,k_2}\delta_{j_2,k_1}+\delta_{i_1,l_2}\delta_{i_2,l_1}\delta_{j_1,k_1}\delta_{j_2,k_2}}{d\left(d^2-1\right)}.
\end{split}
\end{equation}
Furthermore, as we consider the alternating layered ansatz (ALA) in our analysis, we show the five Lemmas derived and shown in Supplementary Information of Ref.~\cite{cerezo2021cost} below.
In these Lemmas, we denote a unitary operator $W$ acting on the Hilbert space $\mathcal{H}_w$ and $W'$ acting on the bipartite system $\mathcal{H}_{w_1}\otimes\mathcal{H}_{w_2}$ as follows.

\begin{equation}
    \label{eq:expr_mat}
    W = \sum_{i,j} W_{i,j}\ket{i}\bra{j}, \quad W'=\sum_{i_1,j_1,i_2,j_2} W'_{i_1 j_1,i_2 j_2} \ket{i_1 i_2}\bra{j_1 j_2}.
\end{equation}

\begin{lemma}
\label{lem1}
Let a unitary $W$ acting on the $d$-dimensional Hilbert space $\mathcal{H}_w$ be a $t$-design with $t\ge1$.
Then, for arbitrary operators $A,B: \mathcal{H}_w\to \mathcal{H}_w$, we have
\begin{equation}
\label{eq:lem1}
    \sum_i p_i \mathrm{Tr}\left[W_i A W_i^{\dagger} B\right] = \int d\mu(W) \mathrm{Tr}\left[WAW^{\dagger}B\right] = \frac{\mathrm{Tr}\left[A\right]\mathrm{Tr}\left[B\right]}{d}.
\end{equation}
\end{lemma}

\begin{lemma}
\label{lem2}
Let a unitary $W$ acting on the $d$-dimensional Hilbert space $\mathcal{H}_w$ be a $t$-design with $t\ge2$.
Then, for arbitrary operators $A,B,C,D: \mathcal{H}_w\to \mathcal{H}_w$, we have

\begin{equation}
\label{eq:lem2}
\begin{split}
    \sum_i p_i \mathrm{Tr}\left[W_i A W_i^{\dagger} B W_i C W_i^{\dagger} D\right] &= \int d\mu(W) \mathrm{Tr}\left[WAW^{\dagger}BWCW^{\dagger}D\right]\\
    &= \frac{1}{d^2-1}\left(\mathrm{Tr}\left[A\right]\mathrm{Tr}\left[C\right]\mathrm{Tr}\left[BD\right]+\mathrm{Tr}\left[AC\right]\mathrm{Tr}\left[B\right]\mathrm{Tr}\left[D\right]\right)\\
    & \qquad -\frac{1}{d\left(d^2-1\right)}\left(\mathrm{Tr}\left[A\right]\mathrm{Tr}\left[B\right]\mathrm{Tr}\left[C\right]\mathrm{Tr}\left[D\right]+\mathrm{Tr}\left[AC\right]\mathrm{Tr}\left[BD\right]\right).
\end{split}
\end{equation}
\end{lemma}

\begin{lemma}
\label{lem3}
Let a unitary $W$ on the $d$-dimensional Hilbert space $\mathcal{H}_w$ be a $t$-design with $t\ge2$.
Then, for arbitrary operators $A,B,C,D: \mathcal{H}_w\to \mathcal{H}_w$, we have
\begin{equation}
\label{eq:lem3}
\begin{split}
    \sum_i p_i \mathrm{Tr}\left[W_i A W_i^{\dagger} B\right]\mathrm{Tr}\left[ W_i C W_i^{\dagger} D\right] &= \int d\mu(W) \mathrm{Tr}\left[WAW^{\dagger}B\right]\mathrm{Tr}\left[WCW^{\dagger}D\right]\\
    &= \frac{1}{d^2-1}\left(\mathrm{Tr}\left[A\right]\mathrm{Tr}\left[B\right]\mathrm{Tr}\left[C\right]\mathrm{Tr}\left[D\right]+\mathrm{Tr}\left[AC\right]\mathrm{Tr}\left[BD\right]\right)\\
    & \qquad -\frac{1}{d\left(d^2-1\right)}\left(\mathrm{Tr}\left[A\right]\mathrm{Tr}\left[C\right]\mathrm{Tr}\left[BD\right]+\mathrm{Tr}\left[AC\right]\mathrm{Tr}\left[B\right]\mathrm{Tr}\left[D\right]\right).
\end{split}
\end{equation}
\end{lemma}

\begin{lemma}
\label{lem4}
Let a unitary $W$ acting on the $d_{w}$-dimensional Hilbert space $\mathcal{H}_w$ be a $t$-design with $t\ge2$.
In addition, suppose $\mathcal{H}=\mathcal{H}_{\bar{w}}\otimes \mathcal{H}_{w}$ be $d_{w}d_{\bar{w}}$-dimensional. 
Then, for arbitrary operators $A,B: \mathcal{H}\to \mathcal{H}$, we have
\begin{equation}
\label{eq:lem4_1}
    \sum_i p_i (\mathbb{I}_{\bar{w}}\otimes W_i) A (\mathbb{I}_{\bar{w}}\otimes W^{\dagger}) B = \int d\mu(W) (\mathbb{I}_{\bar{w}}\otimes W) A (\mathbb{I}_{\bar{w}}\otimes W^{\dagger}) B = \frac{\mathrm{Tr}_{w}\left[A\right]\otimes\mathbb{I}_{w}}{d_{w}}B,
\end{equation}
and
\begin{equation}
\label{eq:lem4_2}
    \sum_i p_i \mathrm{Tr}\left[(\mathbb{I}_{\bar{w}}\otimes W_i) A (\mathbb{I}_{\bar{w}}\otimes W^{\dagger}) B\right] = \int d\mu(W) \mathrm{Tr}\left[(\mathbb{I}_{\bar{w}}\otimes W) A (\mathbb{I}_{\bar{w}}\otimes W^{\dagger}) B\right] = \frac{1}{d_{w}}\mathrm{Tr}\left[\mathrm{Tr}_{w}\left[A\right]\mathrm{Tr}_{w}\left[B\right]\right].
\end{equation}
Here, $\mathbb{I}_{w}$($\mathbb{I}_{\bar{w}}$) represents the identity matrix acting on the Hilbert space $\mathcal{H}_{w}$($\mathcal{H}_{\bar{w}}$) and the partial trace over $\mathcal{H}_{w}$($\mathcal{H}_{\bar{w}}$) is denoted as $\mathrm{Tr}_{w}$($\mathrm{Tr}_{\bar{w}}$). 
Also $\bar{A}$ denotes the complement of $A$. 
\end{lemma}

\begin{lemma}
\label{lem5}
Let W be a unitary operator acting on the $d_{w}$-dimensional Hilbert space $\mathcal{H}_w$.
In addition, suppose $\mathcal{H}=\mathcal{H}_{\bar{w}}\otimes \mathcal{H}_{w}$ be $d_{w}d_{\bar{w}}$-dimensional with $d_{w}=2^m$ and $d_{\bar{w}}=2^{n-m}$. 
Then, for arbitrary operators $A,B: \mathcal{H}\to \mathcal{H}$, we have
\begin{equation}
\label{eq:lem5}
    \mathrm{Tr}\left[(\mathbb{I}_{\bar{w}}\otimes W) A (\mathbb{I}_{\bar{w}}\otimes W^{\dagger}) B\right] = \sum_{\bm{p,q}} \mathrm{Tr} \left[WA_{\bm{qp}},W^{\dagger}B_{\bm{pq}}\right],
\end{equation}
where 
\begin{equation}
    A_{\bm{qp}} = \mathrm{Tr}_{\bar{w}}\left[\left(\ket{\bm{p}}\bra{\bm{q}}\otimes\mathbb{I}_{w}\right)A\right],\quad B_{\bm{pq}} = \mathrm{Tr}_{\bar{w}}\left[\left(\ket{\bm{q}}\bra{\bm{p}}\otimes\mathbb{I}_{w}\right)B\right].
\end{equation}
Here $\bm{q}$ and $\bm{p}$ represent bit-strings of length $n-m$.
\end{lemma}

\section{Vanishing similarity issue in projected quantum kernels}
\label{ap: analytical calc}
In this section, we analytically derive expectation value and variance of PQKs for two types of quantum circuits, i.e., globally random quantum circuits acting on all $n$-qubits and alternating layered ansatz (ALA).
The PQK we consider in our analysis is as follows \cite{huang2021power};
\begin{equation} \label{eq:def_pqk_re}
    k_{PQ}^{(\kappa)}(\bm{x},\bm{x'}) = \mathrm{Tr}\left[ \mathrm{Tr}_{\bar{S}_{\kappa}}\left[\rho(\bm{x},\bm{\theta}) \right] \mathrm{Tr}_{\bar{S}_{\kappa}}\left[\rho(\bm{x'},\bm{\theta}) \right]\right]
\end{equation}
where $\rho(\bm{x},\bm{\theta})=U(\bm{x},\bm{\theta})\rho_{0}U^{\dagger}(\bm{x},\bm{\theta})$ with initial state $\rho_{0}$ and the input- and parameter-dependent unitary operator $U(\bm{x},\bm{\theta})$.
$\mathrm{Tr}_{\bar{S}_{\kappa}}\left[ \cdot \right]$ is the partial trace over the subspace $\bar{S}_{\kappa}$.
Also the number of $\kappa$-th qubit(s) is denoted as $n_{\kappa}$.
In our analysis, we assume that $S_{\kappa}$ is completely included in the subspace on which one of the unitary blocks in the last layer of the ALA acts, as is shown in Fig.~\ref{fig:q_circuit} (c). 
We also assume that initial state $\rho_{0}$ is an arbitrary pure state. 
Lastly, we will state the difference of the variance between Eq.~\eqref{eq:def_pqk_re} and the linear PQK in Appendix~\ref{ap: variance_sum}

\subsection{Case (1): globally random quantum circuits}
\label{ap: case_random}
Here we calculate expectation value and variance of the PQK in Eq.~\eqref{eq:def_pqk_re}, considering the $n$-qubit random quantum circuits.

\subsubsection{Expectation value}
We derive expectation value of the PQK.
We assume either $U(\bm{x},\bm{\theta})$ or $U(\bm{x'},\bm{\theta})$ is a $t$-design with $t\ge1$ without loss of generality.
We here utilize the symmetry of the PQK in Eq.~\eqref{eq:def_pqk_re}.
Especially, we assume that $U(\bm{x},\bm{\theta})$ is a $t$-design with $t\ge1$.
Then expectation value of the PQK over the Haar random unitary, $\braket{k_{PQ}^{(\kappa)}}_{ U(\bm{x},\bm{\theta})}$, is calculated as follows;

\begin{equation}
\label{eq:exp_pqk_rqc}
\begin{split}
    \braket{k_{PQ}^{(\kappa)}}_{ U(\bm{x},\bm{\theta})}&=\left\langle\mathrm{Tr}\left[ \mathrm{Tr}_{\bar{S}_{\kappa}}\left[U(\bm{x},\bm{\theta})\rho_{0}U^{\dagger}(\bm{x},\bm{\theta}) \right] \mathrm{Tr}_{\bar{S}_{\kappa}}\left[\rho(\bm{x'},\bm{\theta}) \right]\right]\right\rangle_{ U(\bm{x},\bm{\theta})} \\
    &= \frac{1}{2^{n}} \mathrm{Tr}\left[ \mathrm{Tr}_{\bar{S}_{\kappa}}\left[\mathbb{I} \right] \mathrm{Tr}_{\bar{S}_{\kappa}}\left[\rho(\bm{x'},\bm{\theta}) \right]\right]\\
    &= \frac{2^{n-n_{\kappa}}}{2^{n}} \mathrm{Tr}\left[\rho(\bm{x'},\bm{\theta})\right]\\
    &= \frac{1}{2^{n_{\kappa}}}\\
\end{split}
\end{equation}
where Lemma \ref{lem1} is used for the second equality and the property of the density matrix, i.e., $\mathrm{Tr}\left[\rho\right]=1$, are utilized for the last equality.

\subsubsection{Variance}
Next, we calculate the variance.
The variance ${\rm{Var}}\left[k_{PQ}^{(\kappa)}\right]$ is expressed as ${\rm{Var}}\left[k_{PQ}^{(\kappa)}\right]=\braket{{k_{PQ}^{(\kappa)}}^2}-\braket{k_{PQ}^{(\kappa)}}^2$.
As we have already had $\braket{k_{PQ}^{(\kappa)}}^2=1/2^{2n_{\kappa}}$, we focus on $\braket{{k_{PQ}^{(\kappa)}}^{2}}$.
Here we assume that $U(\bm{x},\bm{\theta})$ and $U(\bm{x'},\bm{\theta})$ are $t$-designs with $t\ge2$. 
Due to the independence of $U(\bm{x},\bm{\theta})$ and $U(\bm{x'},\bm{\theta})$ from our assumptions, expectation value can be obtained by integrating the square of the PQK over these unitaries; that is $\braket{{k_{PQ}^{(\kappa)}}^{2}}=\braket{{k_{PQ}^{(\kappa)}}^{2}}_{U(\bm{x},\bm{\theta}),U'(\bm{x},\bm{\theta})}$.
Thus, we first calculate the expectation value over $U(\bm{x},\bm{\theta})$, i.e., $\braket{{k_{PQ}^{(\kappa)}}^2}_{U(\bm{x},\bm{\theta})}$.
Then we obtain 

\begin{equation}
\label{eq:var_pqk_rqc_1}
\begin{split}
    &\braket{{k_{PQ}^{(\kappa)}}^{2}}_{U(\bm{x},\bm{\theta})}\\&=\left\langle\mathrm{Tr}\left[ \mathrm{Tr}_{\bar{S}_{\kappa}}\left[U(\bm{x},\bm{\theta})\rho_{0}U^{\dagger}(\bm{x},\bm{\theta}) \right] \mathrm{Tr}_{\bar{S}_{\kappa}}\left[\rho(\bm{x'},\bm{\theta}) \right]\right]\mathrm{Tr}\left[ \mathrm{Tr}_{\bar{S}_{\kappa}}\left[U(\bm{x},\bm{\theta})\rho_{0}U^{\dagger}(\bm{x},\bm{\theta}) \right] \mathrm{Tr}_{\bar{S}_{\kappa}}\left[\rho(\bm{x'},\bm{\theta}) \right]\right]\right\rangle_{U(\bm{x},\bm{\theta})}\\
    &=\biggl\langle\mathrm{Tr}\left[ \left(U(\bm{x},\bm{\theta})\rho_{0}U^{\dagger}(\bm{x},\bm{\theta})\otimes\rho(\bm{x'},\bm{\theta})\right) Swap_{S_{\kappa_1},S_{\kappa_2}}\otimes \mathbb{I}_{\bar{S}_{\kappa_1}\otimes\bar{S}_{\kappa_2}}  \right]\\
    & \qquad \qquad \qquad \times \mathrm{Tr}\left[ \left(U(\bm{x},\bm{\theta})\rho_{0}U^{\dagger}(\bm{x},\bm{\theta})\otimes\rho(\bm{x'},\bm{\theta})\right) Swap_{S_{\kappa_1},S_{\kappa_2}}\otimes \mathbb{I}_{\bar{S}_{\kappa_1}\otimes\bar{S}_{\kappa_2}}  \right]\biggr\rangle_{U(\bm{x},\bm{\theta})}\\
    &= \frac{1}{2^{2n}-1}\left(2^{2(n-n_{\kappa})}\mathrm{Tr}\left[\rho_{0}\right]\mathrm{Tr}\left[\rho_{\bm{x'},\bm{\theta}}\right]\mathrm{Tr}\left[\rho_{0}\right]\mathrm{Tr}\left[\rho_{\bm{x'},\bm{\theta}}\right]+2^{n-n_{\kappa}}\mathrm{Tr}\left[\rho_{0}^{2}\right]\mathrm{Tr}\left[\mathrm{Tr}_{\bar{S}_{\kappa}}\left[\rho_{\bm{x'},\bm{\theta}}\right]\mathrm{Tr}_{\bar{S}_{\kappa}}\left[\rho_{\bm{x'},\bm{\theta}}\right]\right]\right)\\
    & \qquad -\frac{1}{2^{n}\left(2^{2n}-1\right)}\Bigl(2^{n-n_{\kappa}}\mathrm{Tr}\left[\rho_{0}\right]\mathrm{Tr}\left[\rho_{0}\right]\mathrm{Tr}\left[\mathrm{Tr}_{\bar{S}_{\kappa}}\left[\rho_{\bm{x'},\bm{\theta}}\right]\mathrm{Tr}_{\bar{S}_{\kappa}}\left[\rho_{\bm{x'},\bm{\theta}}\right]\right]\\
    & \qquad \qquad \qquad \qquad \qquad \qquad \qquad \qquad \qquad \qquad \qquad \qquad \qquad +2^{2(n-n_{\kappa})}\mathrm{Tr}\left[\rho_{0}^{2}\right]\mathrm{Tr}\left[\rho_{\bm{x'},\bm{\theta}}\right]\mathrm{Tr}\left[\rho_{\bm{x'},\bm{\theta}}\right]\Bigr)\\
    &= \frac{2^{n-n_{\kappa}}}{2^{n}\left(2^{n}+1\right)}\left(2^{n-n_{\kappa}}+\mathrm{Tr}\left[\mathrm{Tr}_{\bar{S}_{\kappa}}\left[\rho_{\bm{x'},\bm{\theta}}\right]\mathrm{Tr}_{\bar{S}_{\kappa}}\left[\rho_{\bm{x'},\bm{\theta}}\right]\right]\right).\\
\end{split}
\end{equation}
In the second equality, we utilize the fact that 
\begin{equation}   
\label{eq:equiv_trace_of_ptrace_product}
\mathrm{Tr}\left[\mathrm{Tr}_{\bar{S}}\left[A\right]\mathrm{Tr}_{\bar{S}}\left[B\right]\right] = \mathrm{Tr}\left[\left(A\otimes B\right)Swap_{S_{1},S_{2}}\otimes \mathbb{I}_{\bar{S}_{1}\otimes\bar{S}_{2}} \right]
\end{equation}
where $A,B$ are the arbitrary matrix, $\mathbb{I}_{S_{1}\otimes S_{2}}$ and $Swap_{S_{1}, S_{2}}$ denote the identity operator and the swap operator acting on the systems $S_{1}, S_{2}$, respectively.
Note that the subspace labeled with the number in the subscript (i.e., $i\in\{1,2\}$ in $S_{i}$) dictates one of the duplicated subsystems.
In the third equality, we use the following result
\begin{equation}
\label{eq:lem6}
\begin{split}
&\left\langle\mathrm{Tr}\left[\mathrm{Tr}_{\bar{S}}\left[wAw^{\dagger}\right]\mathrm{Tr}_{\bar{S}}\left[B\right]\right]\mathrm{Tr}\left[\mathrm{Tr}_{\bar{S}}\left[wAw^{\dagger}\right]\mathrm{Tr}_{\bar{S}}\left[B\right]\right]\right\rangle_{w}\\
&= \left\langle\mathrm{Tr}\left[\left(wAw^{\dagger}\otimes B\right)Swap_{S_{1},S_{2}}\otimes \mathbb{I}_{\bar{S}_{1}\otimes\bar{S}_{2}} \right]\mathrm{Tr}\left[\left(wAw^{\dagger}\otimes B\right)Swap_{S_{1},S_{2}}\otimes \mathbb{I}_{\bar{S}_{1}\otimes\bar{S}_{2}} \right]\right\rangle_{w} \\
&= \frac{1}{d^2-1}\left(\left(\frac{d}{dim(S)}\right)^2\mathrm{Tr}\left[A\right]\mathrm{Tr}\left[B\right]\mathrm{Tr}\left[A\right]\mathrm{Tr}\left[B\right]+\frac{d}{dim(S)}\mathrm{Tr}\left[A^{2}\right]\mathrm{Tr}\left[\mathrm{Tr}_{\bar{S}}\left[B\right]\mathrm{Tr}_{\bar{S}}\left[B\right]\right]\right)\\
& \qquad -\frac{1}{d\left(d^{2}-1\right)}\left(\frac{d}{dim(S)}\mathrm{Tr}\left[A\right]\mathrm{Tr}\left[A\right]\mathrm{Tr}\left[\mathrm{Tr}_{\bar{S}}\left[B\right]\mathrm{Tr}_{\bar{S}}\left[B\right]\right]+\left(\frac{d}{dim(S)}\right)^2\mathrm{Tr}\left[A^{2}\right]\mathrm{Tr}\left[B\right]\mathrm{Tr}\left[B\right]\right)\\
\end{split}
\end{equation}
with the arbitrary matrix $A,B$ of size $d$ and a $d\times d$ unitary matrix $w$.
Here $dim(S)$ represents the dimension of the space $S$.
We note that Eq.~\eqref{eq:lem6} can be obtained using Eq.~\eqref{eq:2d_int} and the property of swap operators regarding the trace operation, i.e., $\mathrm{Tr}[Swap_{S_{1},S_{2}}] = dim(S)$ where $dim(S_1)=dim(S_2)=dim(S)$.
Also, in the last equality, the property of the pure state, i.e., $\mathrm{Tr}\left[\rho\right]=\mathrm{Tr}\left[\rho^2\right]=1$ is used.

In Eq.~\eqref{eq:var_pqk_rqc_1}, only the second term in the last equality depends on $U(\bm{x'},\bm{\theta})$.
Then expectation value of the term can be calculated as
\begin{equation}
\begin{split}
&\left\langle\mathrm{Tr}\left[\mathrm{Tr}_{\bar{S}_{\kappa}}\left[\rho_{\bm{x'},\bm{\theta}}\right]\mathrm{Tr}_{\bar{S}_{\kappa}}\left[\rho_{\bm{x'},\bm{\theta}}\right]\right]\right\rangle_{U(\bm{x'},\bm{\theta})}\\
&= \left\langle\mathrm{Tr}\left[\left(U(\bm{x'},\bm{\theta})\rho_{0}U^{\dagger}(\bm{x'},\bm{\theta})\otimes U(\bm{x'},\bm{\theta})\rho_{0}U^{\dagger}(\bm{x'},\bm{\theta})\right)Swap_{S_{\kappa_1},S_{\kappa_2}}\otimes \mathbb{I}_{\bar{S}_{\kappa_1}\otimes\bar{S}_{\kappa_2}} \right] \right\rangle_{U(\bm{x'},\bm{\theta})} \\
&= \frac{1}{2^{2n}-1}\biggl(\mathrm{Tr}\left[\rho_{0}\right]\mathrm{Tr}\left[\rho_{0}\right]\mathrm{Tr}\left[Swap_{S_{\kappa_1},S_{\kappa_2}}\otimes \mathbb{I}_{\bar{S}_{\kappa_1}\otimes\bar{S}_{\kappa_2}}\right] \\
& \qquad \qquad \qquad \qquad \qquad \qquad  +\mathrm{Tr}\left[\rho_{0}^{2}\right]\mathrm{Tr}\left[Swap_{S_{\kappa_1}\cup\bar{S}_{\kappa_1},S_{\kappa_2}\cup\bar{S}_{\kappa_2}}\left(Swap_{S_{\kappa_1},S_{\kappa_2}}\otimes \mathbb{I}_{\bar{S}_{\kappa_1}\otimes\bar{S}_{\kappa_2}}\right)\right]\biggr)\\
& \qquad -\frac{1}{2^{n}\left(2^{2n}-1\right)}\biggl(\mathrm{Tr}\left[\rho_{0}\right]\mathrm{Tr}\left[\rho_{0}\right]\mathrm{Tr}\left[Swap_{S_{\kappa_1}\cup\bar{S}_{\kappa_1},S_{\kappa_2}\cup\bar{S}_{\kappa_2}}\left(Swap_{S_{\kappa_1},S_{\kappa_2}}\otimes \mathbb{I}_{\bar{S}_{\kappa_1}\otimes\bar{S}_{\kappa_2}}\right)\right]\\
& \qquad \qquad \qquad \qquad \qquad \qquad \qquad \qquad \qquad \qquad \qquad \qquad +\mathrm{Tr}\left[\rho_{0}^{2}\right]\mathrm{Tr}\left[Swap_{S_{\kappa_1},S_{\kappa_2}}\otimes \mathbb{I}_{\bar{S}_{\kappa_1}\otimes\bar{S}_{\kappa_2}}\right]\biggr)\\
&= \frac{1}{2^{2n}-1}\left(1-\frac{1}{2^n}\right)\left(2^{2n-n_\kappa}+2^{n+n_\kappa}\right)\\
&=\frac{1}{2^n+1}\left(2^{n-n_\kappa}+2^{n_\kappa}\right)
\end{split}
\end{equation}
where Eqs.~\eqref{eq:equiv_trace_of_ptrace_product} and ~\eqref{eq:2d_int} are used for the first and second equality and the property of the swap operator with respect to the trace operation is utilized for the third equality.
Therefore we can obtain
\begin{equation}
\begin{split}
    \braket{{k_{PQ}^{(\kappa)}}^{2}} = \frac{2^{n-n_{\kappa}}}{2^{n}\left(2^{n}+1\right)}\left(2^{n-n_{\kappa}}+\frac{1}{2^n+1}\left(2^{n-n_\kappa}+2^{n_\kappa}\right)\right)
\end{split}
\end{equation}
As a result, we have
\begin{equation}
\label{eq:var_pqk_rqc}
\begin{split}
    {\rm{Var}}[k_{PQ}^{(\kappa)}] &= \braket{{k_{PQ}^{(\kappa)}}^{2}}-\braket{k_{PQ}^{(\kappa)}}^2\\
    &= \frac{2^{n-n_{\kappa}}}{2^{n}\left(2^{n}+1\right)}\left(2^{n-n_{\kappa}}+\frac{1}{2^n+1}\left(2^{n-n_\kappa}+2^{n_\kappa}\right)\right)- \frac{1}{2^{2n_{\kappa}}}\\
    &= \frac{2^{2n_{\kappa}}-1}{2^{2n_{\kappa}}\left(2^{n}+1\right)^2}.\\
\end{split}
\end{equation}
We note that the same result can be obtained for the case that different initial states are prepared for $\rho(\bm{x},\bm{\theta})$ and $\rho(\bm{x'},\bm{\theta})$.

\subsection{Case (2): alternating layered ansatzes}
\label{ap: case_ala}
In what follows, we calculate expectation value and variance of the PQK in Eq.~\eqref{eq:def_pqk_re} considering the ALA.

\subsubsection{Expectation value}
We notice that expectation value $\braket{k_{PQ}^{(\kappa)}}_{U(\bm{x},\bm{\theta})}$ can be obtained by integrating the quantity over every unitary block, that is, $\braket{k_{PQ}^{(\kappa)}}_{W_{1,1}(\bm{x},\bm{\theta}),W_{2,1}(\bm{x},\bm{\theta}),\ldots W_{\zeta,L}(\bm{x},\bm{\theta}_{\zeta,L})}$.
Thus, we start with the integration over the unitary block in the last layer that act on $\kappa$-th qubit(s), which we denote $\tilde{W}$.
Then, we obtain
\begin{equation}
\label{eq:exp_pqk_ala_pre}
\begin{split}
    \braket{k_{PQ}^{(\kappa)}}_{\tilde{W}}&=\left\langle\mathrm{Tr}\left[ \mathrm{Tr}_{\bar{S}_{\kappa}}\left[\tilde{W}\rho_{\bm{x},r}\tilde{W}^{\dagger} \right] \mathrm{Tr}_{\bar{S}_{\kappa}}\left[\rho(\bm{x'},\bm{\theta}) \right]\right]\right\rangle_{\tilde{W}} \\
    &= \left\langle\mathrm{Tr}\left[\left(\tilde{W}\rho_{\bm{x},r}\tilde{W}^{\dagger}\otimes \rho(\bm{x'},\bm{\theta})\right)Swap_{S_{\kappa_1},S_{\kappa_2}}\otimes \mathbb{I}_{\bar{S}_{\kappa_1}\otimes\bar{S}_{\kappa_2}} \right]\right\rangle_{\tilde{W}} \\
    &= \frac{1}{2^{m}}  \mathrm{Tr}\left[\left(\mathrm{Tr}_{S_{\Tilde{W}}}\left[\rho_{\bm{x},r} \right]\otimes \rho(\bm{x'},\bm{\theta})\right)\mathrm{Tr}_{S_{\tilde{W}}}\left[Swap_{S_{\kappa_1},S_{\kappa_2}}\otimes \mathbb{I}_{\bar{S}_{\kappa_1}\otimes\bar{S}_{\kappa_2}}\right] \right] \\
    &= \frac{1}{2^{m}}  \mathrm{Tr}\left[\left(\mathrm{Tr}_{S_{\Tilde{W}}}\left[\rho_{\bm{x},r} \right]\otimes \rho(\bm{x'},\bm{\theta})\right) \frac{2^m}{2^{n_{\kappa}}} \mathbb{I}\right] \\
    &= \frac{1}{2^{n_{\kappa}}} \mathrm{Tr}\left[\rho_{\bm{x},r}\right]\mathrm{Tr}\left[\rho(\bm{x'},\bm{\theta})\right]\\
    &= \frac{1}{2^{n_\kappa}}\\
\end{split}
\end{equation}
where Lemma \ref{lem4} is used for the third equality, the property of the swap operator for the trace operation is used for the fourth equality and the property of the density matrix, i.e., $\mathrm{Tr}\left[\rho\right]=1$, is utilized for the last equality.
Here, $\rho_{\bm{x},r}$ is the quantum state resulting from the initial state to which the unitary operator $U(\bm{x},\bm{\theta})$ except for $\tilde{W}$ is applied; namely the equality $\rho(\bm{x},\bm{\theta})=\tilde{W}\rho_{\bm{x},r}\tilde{W}^{\dagger}$ holds. 
As is dictated in Eq.~\eqref{eq:exp_pqk_ala_pre}, the rest of the unitary blocks in the ALA does not contribute to the calculation of the expectation value.
Namely, the unitary blocks can be canceled out in terms of the calculation after the integration over $\Tilde{W}$.
Thus the expectation value reads
\begin{equation}
\label{eq:exp_pqk_ala}
    \braket{k_{PQ}^{(\kappa)}}=\braket{k_{PQ}^{(\kappa)}}_{\tilde{W}}=\frac{1}{2^{n_{\kappa}}}
\end{equation}

\subsubsection{Variance}
Lastly, we compute variance of PQK for ALAs.
As variance can be written as ${\rm{Var}}\left[k_{PQ}^{(\kappa)}\right]=\braket{{k_{PQ}^{(\kappa)}}^{2}}-\braket{k_{PQ}^{(\kappa)}}^2$, we again focus on the quantity, $\braket{{k_{PQ}^{(\kappa)}}^{2}}$.
Also, in analogous to the calculation for the expectation value, we integrate the quantity over all local unitary blocks in the ALAs, $U(\bm{x},\bm{\theta})$ and $U(\bm{x'},\bm{\theta})$.
Especially, we begin with the integration over the unitary blocks in the last layer.
Without loss of generality, we assume that the unitary block in the last layer that acts on $\kappa$-th qubit(s) is the $p$-th unitary block in the last layer, i.e., $W_{p,L}(\bm{x},\bm{\theta}_{p,L})$.
Moreover, for the sake of clarity, we denote $W_{l,d}(\bm{x},\bm{\theta}_{l,d})\equiv W_{l,d}$ ($W_{l,d}(\bm{x'},\bm{\theta}_{l,d})\equiv W'_{l,d}$) hereafter.

To obtain expectation value over the unitary blocks in the last layer, we have to calculate the integration of the following quantity repeatedly;
\begin{equation}
\label{eq:var_quantity_pqk}
    \left\langle \mathrm{Tr}\left[\mathrm{Tr}_{\bar{S}}\left[wAw^{\dagger}\right]\mathrm{Tr}_{\bar{S}}\left[B\right]\right]\mathrm{Tr}\left[\mathrm{Tr}_{\bar{S}}\left[wAw^{\dagger}\right]\mathrm{Tr}_{\bar{S}}\left[B\right]\right]\right\rangle_{w}.
\end{equation}
Thus, based on the calculation to be performed below, we consider the following situations: a unitary block $w$ of size $d_{w}$ acting on (1) a subspace of $\bar{S}$ and (2) a subspace of both $S$ and $\bar{S}$.
Then, for arbitrary operators $A,B: S \otimes \bar{S} \to S \otimes \bar{S}$, the expectation value of Eq.~\eqref{eq:var_quantity_pqk} over $w: S_{w} \to S_{w}$ can be obtained as follows;
\begin{enumerate}
    \item $S_{w} \subseteq \bar{S}$ 
    \begin{equation}
    \begin{split} \label{eq:vq_calc1}
        &\left\langle \mathrm{Tr}\left[\mathrm{Tr}_{\bar{S}}\left[wAw^{\dagger}\right]\mathrm{Tr}_{\bar{S}}\left[B\right]\right]\mathrm{Tr}\left[\mathrm{Tr}_{\bar{S}}\left[wAw^{\dagger}\right]\mathrm{Tr}_{\bar{S}}\left[B\right]\right]\right\rangle_{w} \\
        & \qquad = \left\langle \mathrm{Tr}\left[\mathrm{Tr}_{\bar{S}}\left[Aw^{\dagger}w\right]\mathrm{Tr}_{\bar{S}}\left[B\right]\right]\mathrm{Tr}\left[\mathrm{Tr}_{\bar{S}}\left[Aw^{\dagger}w\right]\mathrm{Tr}_{\bar{S}}\left[B\right]\right]\right\rangle_{w}\\
        & \qquad =  \mathrm{Tr}\left[\mathrm{Tr}_{\bar{S}}\left[A\right]\mathrm{Tr}_{\bar{S}}\left[B\right]\right]\mathrm{Tr}\left[\mathrm{Tr}_{\bar{S}}\left[A\right]\mathrm{Tr}_{\bar{S}}\left[B\right]\right]\\
    \end{split}
    \end{equation}
    \item $S_{w} = S\otimes S_{\bar{h}}$ with $S_{\bar{h}} \subset \bar{S}$ 
    \begin{equation}
    \begin{split} \label{eq:vq_calc2} &\left\langle\mathrm{Tr}\left[\mathrm{Tr}_{\bar{S}}\left[wAw^{\dagger}\right]\mathrm{Tr}_{\bar{S}}\left[B\right]\right]\mathrm{Tr}\left[\mathrm{Tr}_{\bar{S}}\left[wAw^{\dagger}\right]\mathrm{Tr}_{\bar{S}}\left[B\right]\right]\right\rangle_{w}\\
    &= \left\langle\mathrm{Tr}\left[\left(wAw^{\dagger}\otimes B\right)Swap_{S_{1},S_{2}}\otimes \mathbb{I}_{\bar{S}_{1}\otimes\bar{S}_{2}} \right]\mathrm{Tr}\left[\left(wAw^{\dagger}\otimes B\right)Swap_{S_{1},S_{2}}\otimes \mathbb{I}_{\bar{S}_{1}\otimes\bar{S}_{2}} \right]\right\rangle_{w} \\
    &= \frac{1}{d_{w}^2-1}\left(\left(\frac{d_{w}}{dim(S)}\right)^2\mathrm{Tr}\left[A\right]\mathrm{Tr}\left[B\right]\mathrm{Tr}\left[A\right]\mathrm{Tr}\left[B\right]+\frac{d_{w}}{dim(S)}\mathrm{Tr}\left[\mathrm{Tr}_{\bar{S}_{w}}\left[A\right]\mathrm{Tr}_{\bar{S}_{w}}\left[A\right]\right]\mathrm{Tr}\left[\mathrm{Tr}_{\bar{S}}\left[B\right]\mathrm{Tr}_{\bar{S}}\left[B\right]\right]\right)\\
    & \qquad -\frac{1}{d_{w}\left(d_{w}^{2}-1\right)}\Biggl(\frac{d_{w}}{dim(S)}\mathrm{Tr}\left[A\right]\mathrm{Tr}\left[A\right]\mathrm{Tr}\left[\mathrm{Tr}_{\bar{S}}\left[B\right]\mathrm{Tr}_{\bar{S}}\left[B\right]\right] \\
    & \qquad \qquad \qquad \qquad \qquad \qquad \qquad \qquad \qquad +\left(\frac{d_{w}}{dim(S)}\right)^2\mathrm{Tr}\left[\mathrm{Tr}_{\bar{S}_{w}}\left[A\right]\mathrm{Tr}_{\bar{S}_{w}}\left[A\right]\right]\mathrm{Tr}\left[B\right]\mathrm{Tr}\left[B\right]\Biggr).\\
    \end{split}
    \end{equation}
\end{enumerate}
Then we can obtain
\begin{equation}
\label{eq:var_pqk_mid1}
\begin{split}
&\braket{{k_{PQ}^{(\kappa)}}^{2}}_{W_{1,L},\ldots,W_{\zeta,L}}\\
&=\left\langle\mathrm{Tr}\left[ \mathrm{Tr}_{\bar{S}_{\kappa}}\left[\rho(\bm{x},\bm{\theta}) \right] \mathrm{Tr}_{\bar{S}_{\kappa}}\left[\rho(\bm{x'},\bm{\theta}) \right]\right]\mathrm{Tr}\left[ \mathrm{Tr}_{\bar{S}_{\kappa}}\left[\rho(\bm{x},\bm{\theta}) \right] \mathrm{Tr}_{\bar{S}_{\kappa}}\left[\rho(\bm{x'},\bm{\theta}) \right]\right]\right\rangle_{W_{1,L},\ldots,W_{\zeta,L}}\\
&=\left\langle\mathrm{Tr}\left[ \mathrm{Tr}_{\bar{S}_{\kappa}}\left[W_{\kappa,L}\rho_{\bm{x},L-1}W_{\kappa,L}^{\dagger} \right] \mathrm{Tr}_{\bar{S}_{\kappa}}\left[\rho(\bm{x'},\bm{\theta}) \right]\right]\mathrm{Tr}\left[ \mathrm{Tr}_{\bar{S}_{\kappa}}\left[W_{\kappa,L}\rho_{\bm{x},L-1}W_{\kappa,L}^{\dagger} \right] \mathrm{Tr}_{\bar{S}_{\kappa}}\left[\rho(\bm{x'},\bm{\theta}) \right]\right]\right\rangle_{W_{p,L}}\\
&= \frac{1}{2^{2m}-1}\Biggl(\left(\frac{2^{m}}{2^{n_{\kappa}}}\right)^2\mathrm{Tr}\left[\rho_{\bm{x},L-1}\right]\mathrm{Tr}\left[\rho(\bm{x'},\bm{\theta})\right]\mathrm{Tr}\left[\rho_{\bm{x},L-1}\right]\mathrm{Tr}\left[\rho(\bm{x'},\bm{\theta})\right]\\
& \qquad \qquad \qquad \qquad+\frac{2^{m}}{2^{n_{\kappa}}}\mathrm{Tr}\left[\mathrm{Tr}_{\bar{S}_{(p,L)}}\left[\rho_{\bm{x},L-1}\right]\mathrm{Tr}_{\bar{S}_{(p,L)}}\left[\rho_{\bm{x},L-1}\right]\right]\mathrm{Tr}\left[\mathrm{Tr}_{\bar{S}_{\kappa}}\left[\rho(\bm{x'},\bm{\theta})\right]\mathrm{Tr}_{\bar{S}_{\kappa}}\left[\rho(\bm{x'},\bm{\theta})\right]\right]\Biggr)\\
& \qquad -\frac{1}{2^{m}\left(2^{2m}-1\right)}\Biggl(\frac{2^{m}}{2^{n_{\kappa}}}\mathrm{Tr}\left[\rho_{\bm{x},L-1}\right]\mathrm{Tr}\left[\rho_{\bm{x},L-1}\right]\mathrm{Tr}\left[\mathrm{Tr}_{\bar{S}_{\kappa}}\left[\rho(\bm{x'},\bm{\theta})\right]\mathrm{Tr}_{\bar{S}_{\kappa}}\left[\rho(\bm{x'},\bm{\theta})\right]\right]\\
& \qquad \qquad \qquad \qquad \qquad +\left(\frac{2^{m}}{2^{n_{\kappa}}}\right)^2\mathrm{Tr}\left[\mathrm{Tr}_{\bar{S}_{(p,L)}}\left[\rho_{\bm{x},L-1}\right]\mathrm{Tr}_{\bar{S}_{(p,L)}}\left[\rho_{\bm{x},L-1}\right]\right]\mathrm{Tr}\left[\rho(\bm{x'},\bm{\theta})\right]\mathrm{Tr}\left[\rho(\bm{x'},\bm{\theta})\right]\Biggr)\\
&= \frac{1}{2^{2m}-1}\Biggl(\left(\frac{2^{m}}{2^{n_{\kappa}}}\right)^2+\frac{2^{m}}{2^{n_{\kappa}}}\mathrm{Tr}\left[\mathrm{Tr}_{\bar{S}_{(p,L)}}\left[\rho_{\bm{x},L-1}\right]\mathrm{Tr}_{\bar{S}_{(p,L)}}\left[\rho_{\bm{x},L-1}\right]\right]\mathrm{Tr}\left[\mathrm{Tr}_{\bar{S}_{\kappa}}\left[\rho(\bm{x'},\bm{\theta})\right]\mathrm{Tr}_{\bar{S}_{\kappa}}\left[\rho(\bm{x'},\bm{\theta})\right]\right]\Biggr)\\
& \qquad -\frac{1}{2^{m}\left(2^{2m}-1\right)}\Biggl(\frac{2^{m}}{2^{n_{\kappa}}}\mathrm{Tr}\left[\mathrm{Tr}_{\bar{S_{\kappa}}}\left[\rho(\bm{x'},\bm{\theta})\right]\mathrm{Tr}_{\bar{S}_{\kappa}}\left[\rho(\bm{x'},\bm{\theta})\right]\right]+\left(\frac{2^{m}}{2^{n_{\kappa}}}\right)^2\mathrm{Tr}\left[\mathrm{Tr}_{\bar{S}_{(p,L)}}\left[\rho_{\bm{x},L-1}\right]\mathrm{Tr}_{\bar{S}_{(p,L)}}\left[\rho_{\bm{x},L-1}\right]\right]\Biggr)\\
&= \frac{1}{(2^{2m}-1)2^{2n_{\kappa}}}\Biggl( \left(2^{m}\mathrm{Tr}\left[\mathrm{Tr}_{\bar{S}_{(p,L)}}\left[\rho_{\bm{x},L-1}\right]\mathrm{Tr}_{\bar{S}_{(p,L)}}\left[\rho_{\bm{x},L-1}\right]\right] -1\right) 2^{n_{\kappa}} \mathrm{Tr}\left[\mathrm{Tr}_{\bar{S_{\kappa}}}\left[\rho(\bm{x'},\bm{\theta})\right]\mathrm{Tr}_{\bar{S}_{\kappa}}\left[\rho(\bm{x'},\bm{\theta})\right]\right] \\
& \qquad \qquad \qquad \qquad \qquad \qquad \qquad \qquad \qquad \qquad \qquad \qquad \qquad + 2^m \mathrm{Tr}\left[\mathrm{Tr}_{\bar{S}_{(p,L)}}\left[\rho_{\bm{x},L-1}\right]\mathrm{Tr}_{\bar{S}_{(p,L)}}\left[\rho_{\bm{x},L-1}\right]\right] -2^{2m}\Biggr), \\
\end{split}
\end{equation}
where Eqs~\eqref{eq:vq_calc1} and ~\eqref{eq:vq_calc2} are used in the second and third equality, respectively, and the trace property of density matrix is used in the last equality.
Also $\rho_{\bm{x},d}$ denotes the quantum state resulting from the initial state to which the unitary operator $U(\bm{x},\bm{\theta})$ except for the unitary blocks from $(d+1)$-th layer through the last layer is applied, i.e., $\rho(\bm{x},\bm{\theta})=(\prod_{l=d}^{L}V_{l}(\bm{x},\bm{\theta}))\rho_{\bm{x},d}(\prod_{l=d}^{L}V_{l}(\bm{x},\bm{\theta}))^{\dagger}$.
We remind that $S_{(l,d)}$ denotes the subspace of qubits on which the unitary block $W_{l,d}$ ($W'_{l,d}$) acts.

Next, we compute the integration of $\braket{{k_{PQ}^{(\kappa)}}^{2}}_{W_{1,L},\ldots,W_{\zeta,L}}$ in Eq.~\eqref{eq:var_pqk_mid1} over the unitary blocks in the last layer of $U(\bm{x'},\bm{\theta})$.
In this case, only $\mathrm{Tr}[\mathrm{Tr}_{\bar{S}}[\rho(\bm{x'},\bm{\theta})]\mathrm{Tr}_{\bar{S}}[\rho(\bm{x'},\bm{\theta})]]$ in Eq.~\eqref{eq:var_pqk_mid1} matters.
To compute the integral of the quantity over the unitary blocks in the last layer, the following situations can be expected: a unitary block $w$ acting on (1) a subspace of $S$, (2) a subspace of $\bar{S}$, (3) a subspace of both $S$ and $\bar{S}$ and (4) $S$ and a subspace of $\bar{S}$.
Then, for arbitrary operator $A: S' \otimes \bar{S'} \to S' \otimes \bar{S'}$, expectation value of $ \mathrm{Tr}[\mathrm{Tr}_{\bar{S}}[wAw^{\dagger}]\mathrm{Tr}_{\bar{S}}[wAw^{\dagger}]]$ over $w: S_{w} \to S_{w}$ can be obtained as follows;
\begin{enumerate}
    \item $S_{w} \subseteq S$ 
    \begin{equation}
    \begin{split} \label{eq:ala_int_vav1}
        \left\langle \mathrm{Tr}\left[\mathrm{Tr}_{\bar{S}}\left[wAw^{\dagger}\right]\mathrm{Tr}_{\bar{S}}\left[wAw^{\dagger}\right]\right]\right\rangle_{w}
        &= \left\langle \mathrm{Tr}\left[w\mathrm{Tr}_{\bar{S}}\left[A\right]w^{\dagger}w\mathrm{Tr}_{\bar{S}}\left[A\right]w^{\dagger}\right]\right\rangle_{w}\\
        &= \mathrm{Tr}\left[\mathrm{Tr}_{\bar{S}}\left[A\right]\mathrm{Tr}_{\bar{S}}\left[A\right]\right]
    \end{split}
    \end{equation}
    \item $S_{w} \subset \bar{S}$ 
    \begin{equation}
    \begin{split} \label{eq:ala_int_vav2}
        \left\langle \mathrm{Tr}\left[\mathrm{Tr}_{\bar{S}}\left[wAw^{\dagger}\right]\mathrm{Tr}_{\bar{S}}\left[wAw^{\dagger}\right]\right]\right\rangle_{w}
        &= \left\langle \mathrm{Tr}\left[\mathrm{Tr}_{\bar{S}}\left[Aw^{\dagger}w\right]\mathrm{Tr}_{\bar{S}}\left[Aw^{\dagger}w\right]\right]\right\rangle_{w}\\
        &= \mathrm{Tr}\left[\mathrm{Tr}_{\bar{S}}\left[A\right]\mathrm{Tr}_{\bar{S}}\left[A\right]\right]
    \end{split}
    \end{equation}
    \item $S_{w} = S_{h}\otimes S_{\bar{h}}$ with $d^{1/2}$-dimensional spaces $S_{h} \subset S$ and $S_{\bar{h}} \subset \bar{S}$ 
    \begin{equation}
    \begin{split} \label{eq:ala_int_vav3}
        &\left\langle \mathrm{Tr}\left[\mathrm{Tr}_{\bar{S}}\left[wAw^{\dagger}\right]\mathrm{Tr}_{\bar{S}}\left[wAw^{\dagger}\right]\right]\right\rangle_{w} \\
        &= \left\langle  \mathrm{Tr}\left[\left(wAw^{\dagger} \otimes wAw^{\dagger} \right)\left( Swap_{S_{1}, S_{2}} \otimes \mathbb{I}_{\bar{S}_{1}\otimes \bar{S}_{2}}\right)\right] \right\rangle_{w} \\
        & = \frac{1}{d^2-1}\Bigl( \mathrm{Tr}\left[\left(\mathbb{I}_{S_{w,1} \otimes S_{w,2}} \otimes \mathrm{Tr}_{S_{w,1}}\left[A\right] \otimes \mathrm{Tr}_{S_{w,2}}\left[A\right] \right)\left( Swap_{S_{1},S_{2}} \otimes \mathbb{I}_{\bar{S}_{1}\otimes \bar{S}_{2}}\right)\right] \\
        & \quad  + \mathrm{Tr}\left[\left(Swap_{S_{w,1}, S_{w,2}} \otimes \mathrm{Tr}_{S_{w}} \otimes \mathrm{Tr}_{S_{w,1}\cup S_{w,2} }\left[A \otimes A \left( Swap_{S_{w,1}, S_{w,2}} \otimes \mathbb{I}_{\bar{S_{w,1}}\otimes \bar{S_{w,2}}}\right)\right] \right)\left( Swap_{S_{1}, S_{2}} \otimes \mathbb{I}_{\bar{S_{1}}\otimes \bar{S_{2}}}\right)\right] \Bigr) \\
        & \quad - \frac{1}{d(d^2-1)}\Bigl( \mathrm{Tr}\left[\left(\mathbb{I}_{S_{w,1} \otimes S_{w,1}} \otimes \mathrm{Tr}_{S_{w,1}\cup S_{w,2} }\left[A \otimes A \left( Swap_{S_{w,1}, S_{w,2}} \otimes \mathbb{I}_{\bar{S_{w,1}}\otimes \bar{S_{w,2}}}\right)\right] \right)\left( Swap_{S_{1}, S_{2}} \otimes \mathbb{I}_{\bar{S}_{1}\otimes \bar{S}_{2}}\right)\right] \\
        & \quad \quad + \mathrm{Tr}\left[\left(Swap_{S_{w,1}, S_{w,2}} \otimes \mathrm{Tr}_{S_{w,1}}\left[A\right] \otimes \mathrm{Tr}_{S_{w,2}}\left[A\right] \right)\left( Swap_{S_{1}, S_{2}} \otimes \mathbb{I}_{\bar{S}_{1}\otimes \bar{S}_{2}}\right)\right] \Bigr) \\
        & = \frac{d^{1/2}}{d+1}\left(\mathrm{Tr}\left[\mathrm{Tr}_{\bar{S}\cup S_{h}}\left[A\right]\mathrm{Tr}_{\bar{S}\cup S_{h}}\left[A\right]\right] + \mathrm{Tr}\left[\mathrm{Tr}_{\bar{S}\backslash S_{\bar{h}}}\left[A\right]\mathrm{Tr}_{\bar{S}\backslash S_{\bar{h}}}\left[A\right]\right] \right)
    \end{split}
    \end{equation}
    \item $S_{w} = S\otimes S_{\bar{h}}$ with $d^{1/2}$-dimensional spaces $S$ and $S_{\bar{h}} \subset \bar{S}$ 
    \begin{equation}
    \begin{split} \label{eq:ala_int_vav4}
        \left\langle \mathrm{Tr}\left[\mathrm{Tr}_{\bar{S}}\left[wAw^{\dagger}\right]\mathrm{Tr}_{\bar{S}}\left[wAw^{\dagger}\right]\right]\right\rangle_{w} &= \left\langle  \mathrm{Tr}\left[\left(wAw^{\dagger} \otimes wAw^{\dagger} \right)\left( Swap_{S_{1}, S_{2}} \otimes \mathbb{I}_{\bar{S}_{1}\otimes \bar{S}_{2}}\right)\right] \right\rangle_{w} \\
        & = \frac{d^{1/2}}{d+1}\left(\mathrm{Tr}\left[A\right]\mathrm{Tr}\left[A\right] + \mathrm{Tr}\left[\mathrm{Tr}_{\bar{S}\backslash S_{\bar{h}}}\left[A\right]\mathrm{Tr}_{\bar{S}\backslash  S_{\bar{h}}}\left[A\right]\right] \right).
    \end{split}
    \end{equation}
\end{enumerate}

Hence, using Eqs.~\eqref{eq:ala_int_vav1} to ~\eqref{eq:ala_int_vav4}, we obtain
\begin{equation}
\label{eq:var_pqk_mid2}
\begin{split}
    &\left\langle\mathrm{Tr}\left[\mathrm{Tr}_{\bar{S}_{\kappa}}\left[\rho(\bm{x'},\bm{\theta})\right]\mathrm{Tr}_{\bar{S}_{\kappa}}\left[\rho(\bm{x'},\bm{\theta})\right]\right]\right\rangle_{W'_{1,L},\ldots,W'_{\zeta,L}} \\
    &= \left\langle\mathrm{Tr}\left[\mathrm{Tr}_{\bar{S}_{\kappa}}\left[W'_{p,L}\rho_{\bm{x'},L-1}W_{p,L}^{'\dagger}\right]\mathrm{Tr}_{\bar{S}_{\kappa}}\left[W'_{p,L}\rho_{\bm{x'},L-1}W_{p,L}^{'\dagger}\right]\right]\right\rangle_{W'_{p,L}} \\
    &= \frac{1}{2^{2m}-1}\Biggl(\left(\frac{2^{m}}{2^{n_{\kappa}}}\right)^2 2^{n_{\kappa}}+\left(\frac{2^{m}}{2^{n_{\kappa}}}\right) 2^{2n_{\kappa}}\mathrm{Tr}\left[\mathrm{Tr}_{\bar{S}_{(p,L)}}\left[\rho_{\bm{x'},L-1}\right]\mathrm{Tr}_{\bar{S}_{(p,L)}}\left[\rho_{\bm{x'},L-1}\right]\right]\Biggr)\\
    & \qquad -\frac{1}{2^{m}\left(2^{2m}-1\right)}\Biggl(\left(\frac{2^{m}}{2^{n_{\kappa}}}\right)2^{2n_{\kappa}}+\left(\frac{2^{m}}{2^{n_{\kappa}}}\right)^2 2^{n_{\kappa}}\mathrm{Tr}\left[\mathrm{Tr}_{\bar{S}_{(p,L)}}\left[\rho_{\bm{x'},L-1}\right]\mathrm{Tr}_{\bar{S}_{(p,L)}}\left[\rho_{\bm{x'},L-1}\right]\right]\Biggr),\\
    &= \frac{1}{2^{2m}-1} \left(\left(2^{m+n_{\kappa}}-\frac{2^{m}}{2^{n_{\kappa}}}\right)\mathrm{Tr}\left[\mathrm{Tr}_{\bar{S}_{(p,L)}}\left[\rho_{\bm{x'},L-1}\right]\mathrm{Tr}_{\bar{S}_{(p,L)}}\left[\rho_{\bm{x'},L-1}\right]\right] + \frac{2^{2m}}{2^{n_{\kappa}}}-2^{2n_{\kappa}}\right).
\end{split}
\end{equation}

Thus, from Eqs.~\eqref{eq:var_pqk_mid1} and ~\eqref{eq:var_pqk_mid2}, the expectation value $\braket{{k_{PQ}^{(\kappa)}}^2}_{W_{1,L},\ldots,W_{\zeta,L}, W'_{1,L},\ldots,W'_{\zeta,L}}$ can read
\begin{equation}
\begin{split}
   &\braket{{k_{PQ}^{(\kappa)}}^2}_{W_{1,L},\ldots,W_{\zeta,L}, W'_{1,L},\ldots,W'_{\zeta,L}} \\
   &= \frac{1}{(2^{2m}-1)2^{2n_{\kappa}}}\Biggl( \left(2^{m}\mathrm{Tr}\left[\mathrm{Tr}_{\bar{S}_{(p,L)}}\left[\rho_{\bm{x},L-1}\right]\mathrm{Tr}_{\bar{S}_{(p,L)}}\left[\rho_{\bm{x},L-1}\right]\right] -1\right) \\
   &  \qquad \qquad \qquad \qquad \times 2^{n_{\kappa}}  \frac{1}{2^{2m}-1} \left(\left(2^{m+n_{\kappa}}-\frac{2^{m}}{2^{n_{\kappa}}}\right)\mathrm{Tr}\left[\mathrm{Tr}_{\bar{S}_{(p,L)}}\left[\rho_{\bm{x'},L-1}\right]\mathrm{Tr}_{\bar{S}_{(p,L)}}\left[\rho_{\bm{x'},L-1}\right]\right] + \frac{2^{2m}}{2^{n_{\kappa}}}-2^{2n_{\kappa}}\right)\\
    &  \quad \qquad \qquad \qquad \qquad \qquad \qquad \qquad \qquad \qquad \qquad \qquad \quad + 2^m \mathrm{Tr}\left[\mathrm{Tr}_{\bar{S}_{(p,L)}}\left[\rho_{\bm{x},L-1}\right]\mathrm{Tr}_{\bar{S}_{(p,L)}}\left[\rho_{\bm{x},L-1}\right]\right] -2^{2m}\Biggr)\\
    &= \frac{1}{2^{2n_{\kappa}}} + \frac{2^{2m}\left(2^{2n_{\kappa}}-1\right)}{(2^{2m}-1)^{2} 2^{2n_{\kappa}}}\\
    & \qquad \qquad \times \left(\mathrm{Tr}\left[\mathrm{Tr}_{\bar{S}_{(p,L)}}\left[\rho_{\bm{x},L-1}\right]\mathrm{Tr}_{\bar{S}_{(p,L)}}\left[\rho_{\bm{x},L-1}\right]\right]-\frac{1}{2^{m}}\right)\left(\mathrm{Tr}\left[\mathrm{Tr}_{\bar{S}_{(p,L)}}\left[\rho_{\bm{x'},L-1}\right]\mathrm{Tr}_{\bar{S}_{(p,L)}}\left[\rho_{\bm{x'},L-1}\right]\right]-\frac{1}{2^{m}}\right) \\
    &= \frac{1}{2^{2n_{\kappa}}} + \frac{2^{2m}\left(2^{2n_{\kappa}}-1\right)}{(2^{2m}-1)^{2} 2^{2n_{\kappa}}} \left(\mathrm{Tr}\left[\mathrm{Tr}_{\bar{S}_{(p,L)}}\left[\rho_{\bm{x},L-1}\right]\mathrm{Tr}_{\bar{S}_{(p,L)}}\left[\rho_{\bm{x},L-1}\right]\right]-\frac{1}{2^{m}}\right)^2
\end{split}
\end{equation}
In the last equality, we utilize our assumption that any unitary block in $U(\bm{x},\bm{\theta})$ and $U(\bm{x'},\bm{\theta})$ are $2$-designs, and an additional assumption that the same initial state is prepared for $\rho(\bm{x},\bm{\theta})$ and $\rho(\bm{x'},\bm{\theta})$.
Therefore, variance of PQK for ALAs can be written as
\begin{equation}
\label{eq:var_pqk_ala_mid}
\begin{split}
    {\rm{Var}}[k_{PQ}^{(\kappa)}] &= \braket{{k_{PQ}^{(\kappa)}}^{2}}-\braket{k_{PQ}^{(\kappa)}}^2\\
    &= \left\langle\braket{{k_{PQ}^{(\kappa)}}^2}_{W_{1,L},\ldots,W_{\zeta,L}, W'_{1,L},\ldots,W'_{\zeta,L}} \right\rangle_{W_{1,1},\ldots,W_{\zeta,L-1},W'_{1,1},\ldots,W'_{\zeta,L-1}} -\braket{k_{PQ}^{(\kappa)}}^2\\
    &= \frac{1}{2^{2n_{\kappa}}} + \frac{2^{2m}\left(2^{2n_{\kappa}}-1\right)}{(2^{2m}-1)^{2} 2^{2n_{\kappa}}} \left(\left\langle\mathrm{Tr}\left[\mathrm{Tr}_{\bar{S}_{(p,L)}}\left[\rho_{\bm{x},L-1}\right]\mathrm{Tr}_{\bar{S}_{(p,L)}}\left[\rho_{\bm{x},L-1}\right]\right]\right\rangle_{W_{1,1},\ldots,W_{\zeta,L-1}}-\frac{1}{2^{m}}\right)^2- \frac{1}{2^{2n_{\kappa}}}\\
    &= \frac{2^{2m}\left(2^{2n_{\kappa}}-1\right)}{(2^{2m}-1)^{2} 2^{2n_{\kappa}}} \left(\left\langle\mathrm{Tr}\left[\mathrm{Tr}_{\bar{S}_{(p,L)}}\left[\rho_{\bm{x},L-1}\right]\mathrm{Tr}_{\bar{S}_{(p,L)}}\left[\rho_{\bm{x},L-1}\right]\right]\right\rangle_{W_{1,1},\ldots,W_{\zeta,L-1}}-\frac{1}{2^{m}}\right)^2.\\
\end{split}
\end{equation}

The implication of Eq.~\eqref{eq:var_pqk_ala_mid} is that the variance depends on the purity of the quantum state, i.e., $\mathrm{Tr}[(\rho_{\bm{x},L-1}^{S_{(p,L)}})^2]$ with $\rho_{\bm{x},L-1}^{S_{(p,L)}} \equiv \mathrm{Tr}_{\bar{S}_{(p,L)}}\left[\rho_{\bm{x},L-1}\right]$.
We remind that $\rho_{\bm{x},L-1}$ is the quantum state resulting from the initial state to which the unitary operator $U(\bm{x},\bm{\theta})$ except for the unitary blocks in the last layer is applied, and $\mathrm{Tr}_{\bar{S}_{(p,L)}}\left[\cdot\right]$ is the partial trace over the subspace $\bar{S}_{(p,L)}$ with $S_{(p,L)}$ the subspace on which the unitary block $W_{p,L}$ acts.
This means that the variance is zero if $\rho_{\bm{x},L-1}^{S_{(p,L)}}$ is the completely mixed state, i.e., $\mathbb{I}/2^{m}$. 
Thus, Eq.~\eqref{eq:var_pqk_ala_mid} indicates that how fast the quantum state converges to the mixed state is of crucial to avoid the trainability issue, while the situation where $\rho_{\bm{x},L-1}^{S_{(p,L)}}$ is close to a pure state means the unitary operation is efficiently simulatable by classical computers. 

Lastly, we check the relationship between the variance and initial state as well as circuit depth of ALA.
We here consider two cases regarding the position of the $\kappa$-th qubit(s): $\kappa$-th qubit(s) is (are) located (1) in the middle so that the number of qubits on which the unitary blocks in the first layer inside the light-cone act is smaller than $n$ and (2) in the unitary block at the edge i.e., $W_{1,L}$ or $W_{\zeta,L}$.
Case (1) corresponds to the situation where the the number of unitary blocks inside the light-cone is the smallest and case (2) is the one where the number of blocks is the largest.

For ease of understanding, we first consider one-layer ALAs.
In this case, the variance for both case (1) and (2) can be written as
\begin{equation}
\label{eq:var_one_layer}
    {\rm{Var}}[k_{PQ}^{(\kappa)}] =  \frac{2^{2m}\left(2^{2n_{\kappa}}-1\right)}{(2^{2m}-1)^{2} 2^{2n_{\kappa}}} \left(\mathrm{Tr}\left[\mathrm{Tr}_{\bar{S}_{(p,1)}}\left[\rho_0\right]\mathrm{Tr}_{\bar{S}_{(p,1)}}\left[\rho_0\right]\right]-\frac{1}{2^{m}}\right)^2.
\end{equation}
Next, we deal with two-layer ALAs.
Then, using Eqs.~\eqref{eq:ala_int_vav1} to ~\eqref{eq:ala_int_vav4}, we obtain 
\begin{equation}
\label{eq:var_two_layer}
\begin{split}
    {\rm{Var}}[k_{PQ}^{(\kappa)}] &=
    \frac{2^{2m}\left(2^{2n_{\kappa}}-1\right)}{(2^{2m}-1)^{2} 2^{2n_{\kappa}}} \left(\left\langle\mathrm{Tr}\left[\mathrm{Tr}_{\bar{S}_{(p,2)}}\left[\rho_{\bm{x},1}\right]\mathrm{Tr}_{\bar{S}_{(p,2)}}\left[\rho_{\bm{x},1}\right]\right]\right\rangle_{W_{1,1},\ldots,W_{\zeta,1}}-\frac{1}{2^{m}}\right)^2\\
    &= \frac{2^{2m}\left(2^{2n_{\kappa}}-1\right)}{(2^{2m}-1)^{2} 2^{2n_{\kappa}}} \biggl(\frac{2^{m}}{(2^{m}+1)^2}\Bigl(1+\mathrm{Tr}\left[\mathrm{Tr}_{\bar{S}_{(p,1)}}\left[\rho_0\right]\mathrm{Tr}_{\bar{S}_{(p,1)}}\left[\rho_0\right]\right]\\
    & \qquad \qquad  +\mathrm{Tr}\left[\mathrm{Tr}_{\bar{S}_{(p+1,1)}}\left[\rho_0\right]\mathrm{Tr}_{\bar{S}_{(p+1,1)}}\left[\rho_0\right]\right]+\mathrm{Tr}\left[\mathrm{Tr}_{\overline{S_{(p,1)}\cup S_{ (p+1,1)}}}\left[\rho_0\right]\mathrm{Tr}_{\overline{S_{(p,1)}\cup S_{ (p+1,1)}}}\left[\rho_0\right]\right]\Bigr)-\frac{1}{2^{m}}\biggr)^2 \\
    &= \frac{2^{2m}\left(2^{2n_{\kappa}}-1\right)}{(2^{2m}-1)^{2}(2^m+1)^4 2^{2n_{\kappa}}}\biggl(-2-\frac{1}{2^{m}}+2^{m}\mathrm{Tr}\left[\mathrm{Tr}_{\bar{S}_{(p,1)}}\left[\rho_0\right]\mathrm{Tr}_{\bar{S}_{(p,1)}}\left[\rho_0\right]\right]\\
    & \qquad \qquad  +2^{m}\mathrm{Tr}\left[\mathrm{Tr}_{\bar{S}_{(p+1,1)}}\left[\rho_0\right]\mathrm{Tr}_{\bar{S}_{(p+1,1)}}\left[\rho_0\right]\right]+2^{m}\mathrm{Tr}\left[\mathrm{Tr}_{\overline{S_{(p,1)}\cup S_{ (p+1,1)}}}\left[\rho_0\right]\mathrm{Tr}_{\overline{S_{(p,1)}\cup S_{ (p+1,1)}}}\left[\rho_0\right]\right]\biggr)^2
\end{split}
\end{equation}
for case (1) and (2).

Subsequently, for case (1) with a three-layer ALA, we have
\begin{equation}
\label{eq:var_three_layer}
\begin{split}
    {\rm{Var}}[k_{PQ}^{(\kappa)}] &=
    \frac{2^{2m}\left(2^{2n_{\kappa}}-1\right)}{(2^{2m}-1)^{2} 2^{2n_{\kappa}}} \left(\left\langle\mathrm{Tr}\left[\mathrm{Tr}_{\bar{S}_{(p,3)}}\left[\rho_{\bm{x},2}\right]\mathrm{Tr}_{\bar{S}_{(p,3)}}\left[\rho_{\bm{x},2}\right]\right]\right\rangle_{W_{1,1},\ldots,W_{\zeta,2}}-\frac{1}{2^{m}}\right)^2\\
    &= \frac{2^{4m}\left(2^{2n_{\kappa}}-1\right)}{(2^{2m}-1)^{2}(2^m+1)^8 2^{2n_{\kappa}}}\biggl(-5-\frac{4}{2^{m}}-\frac{1}{2^{2m}} \\
    & \qquad + 2^{m}\Bigl(\mathrm{Tr}\left[\mathrm{Tr}_{\bar{S}_{(p-1,1)}}\left[\rho_0\right]\mathrm{Tr}_{\bar{S}_{(p-1,1)}}\left[\rho_0\right]\right] +\mathrm{Tr}\left[\mathrm{Tr}_{\bar{S}_{(p,1)}}\left[\rho_0\right]\mathrm{Tr}_{\bar{S}_{(p,1)}}\left[\rho_0\right]\right]\\
    & \qquad \qquad \qquad +\mathrm{Tr}\left[\mathrm{Tr}_{\overline{S_{(p-1,1)}\cup S_{ (p,1)}}}\left[\rho_0\right]\mathrm{Tr}_{\overline{S_{(p-1,1)}\cup S_{ (p,1)}}}\left[\rho_0\right]\right]\Bigr) \\
    & \qquad + 2^{m}\Bigl(\mathrm{Tr}\left[\mathrm{Tr}_{\bar{S}_{(p,1)}}\left[\rho_0\right]\mathrm{Tr}_{\bar{S}_{(p,1)}}\left[\rho_0\right]\right] +\mathrm{Tr}\left[\mathrm{Tr}_{\bar{S}_{(p+1,1)}}\left[\rho_0\right]\mathrm{Tr}_{\bar{S}_{(p+1,1)}}\left[\rho_0\right]\right]\\
    & \qquad \qquad \qquad +\mathrm{Tr}\left[\mathrm{Tr}_{\overline{S_{(p,1)}\cup S_{ (p+1,1)}}}\left[\rho_0\right]\mathrm{Tr}_{\overline{S_{(p,1)}\cup S_{ (p+1,1)}}}\left[\rho_0\right]\right]\Bigr) \\
    & \qquad + 2^{m} \Bigl(\mathrm{Tr}\left[\mathrm{Tr}_{\bar{S}_{(p,1)}}\left[\rho_0\right]\mathrm{Tr}_{\bar{S}_{(p,1)}}\left[\rho_0\right]\right]+\mathrm{Tr}\left[\mathrm{Tr}_{\overline{S_{(p,1)}\cup S_{ (p+1,1)}}}\left[\rho_0\right]\mathrm{Tr}_{\overline{S_{(p,1)}\cup S_{ (p+1,1)}}}\left[\rho_0\right]\right]\\
    & \qquad \qquad \qquad +\mathrm{Tr}\left[\mathrm{Tr}_{\overline{S_{(p-1,1)}\cup S_{ (p,1)}}}\left[\rho_0\right]\mathrm{Tr}_{\overline{S_{(p-1,1)}\cup S_{ (p,1)}}}\left[\rho_0\right]\right]\\
    & \qquad \qquad \qquad +\mathrm{Tr}\left[\mathrm{Tr}_{\overline{S_{(p-1,1)}\cup S_{(p,1)}\cup S_{ (p+1,1)}}}\left[\rho_0\right]\mathrm{Tr}_{\overline{S_{(p-1,1)}\cup S_{(p,1)}\cup S_{ (p+1,1)}}}\left[\rho_0\right]\right]\Bigr)\biggr).
\end{split}
\end{equation}
For case (2), we get
\begin{equation}
\label{eq:var_two_layer_edge}
\begin{split}
    {\rm{Var}}[k_{PQ}^{(\kappa)}] 
    &= \frac{2^{2m}\left(2^{2n_{\kappa}}-1\right)}{(2^{2m}-1)^{2}(2^{m}+1)^4 2^{2n_{\kappa}}} \biggl(-1-\frac{1}{2^{m}} +2^{m/2}\Bigl(2\mathrm{Tr}\left[\mathrm{Tr}_{\overline{S_{(p,3)}\backslash S_{ (p,2)}}}\left[\rho_0\right]\mathrm{Tr}_{\overline{S_{(p,3)}\backslash S_{ (p,2)}}}\left[\rho_0\right]\right]\\
    & \qquad \qquad \qquad \qquad \qquad \qquad \qquad \qquad \qquad +\mathrm{Tr}\left[\mathrm{Tr}_{\overline{S_{(p,3)}\cup S_{ (p,2)}}}\left[\rho_0\right]\mathrm{Tr}_{\overline{S_{(p,3)}\cup S_{ (p,2)}}}\left[\rho_0\right]\right]\Bigr)\biggr)^2 \\
    &= \frac{2^{3m}\left(2^{2n_{\kappa}}-1\right)}{(2^{2m}-1)^{2}(2^{m}+1)^4 2^{2n_{\kappa}}} \biggl(-\frac{2}{2^{m/2}}-\frac{1}{2^{3m/2}} +2^{m/2}\Bigl(2\mathrm{Tr}\left[\mathrm{Tr}_{\bar{S}_{(p,1)}}\left[\rho_0\right]\mathrm{Tr}_{\bar{S}_{(p,1)}}\left[\rho_0\right]\right]\\
    & \qquad \qquad \qquad \qquad \qquad \qquad \qquad \qquad +\mathrm{Tr}\left[\mathrm{Tr}_{\overline{S_{(p,1)}\cup S_{ (p+1,1)}}}\left[\rho_0\right]\mathrm{Tr}_{\overline{S_{(p,1)}\cup S_{ (p+1,1)}}}\left[\rho_0\right]\right]\Bigr)\biggr)^2. \\
\end{split}
\end{equation}
We note that we here consider $p=1$ without loss of generality.

Therefore, the variance in case (1) reads
\begin{equation}
\label{eq:var_L_layer}
\begin{split}
    {\rm{Var}}[k_{PQ}^{(\kappa)}] &=
    \frac{2^{2m(L-1)}\left(2^{2n_{\kappa}}-1\right)}{(2^{2m}-1)^{2}(2^m+1)^{4(L-1)} 2^{2n_{\kappa}}}\left(2^{m} \sum_{h\in P(S_{(k_u,1)}:S_{(k_l,1)})} t_{h} \mathrm{Tr}\left[\mathrm{Tr}_{\bar{h}}\left[\rho_0\right]\mathrm{Tr}_{\bar{h}}\left[\rho_0\right]\right] -\sum_{\tau=0}^{L-1} \frac{c_\tau}{2^{m\tau}}, \right)^2, \\
\end{split}
\end{equation}
where $c_{\tau},t_h \in \mathbb{R}^{+} $ and $P(S_{(k_u,1)}:S_{(k_l,1)})=\left\{\bigcup_{i=\xi}^{\xi+l} S_{(k_u+i,1)}| l\in\{1,\ldots,L\}, \xi\in\{0,\ldots,(k_l-k_u)-i+1\}\right\}$ is the set containing all the possible neighboring subspaces in $\bigcup_{i=0}^{k_l-k_u} S_{(k_u+i,1)}$. 
We here define $W_{k_{u},1}$ and $W_{k_{l},1}$ as the the unitary blocks located at the edge of the light-cone in the first layer.
We also remind that $S_{(l,d)}$ denotes the subspace of the qubits that the unitary operator $W_{l,d}$ acts on.\\
As for case (2) with odd $L$ layers ($L\ge3$),
we get
\begin{equation}
\label{eq:var_L_layer_edge_odd}
\begin{split}
    {\rm{Var}}[k_{PQ}^{(\kappa)}] &=
    \frac{2^{mL}\left(2^{2n_{\kappa}}-1\right)}{(2^{2m}-1)^{2}(2^m+1)^{2(L-1)} 2^{2n_{\kappa}}}\left(2^{m/2} \sum_{h\in P'(S_{(k_u,1)}:S_{(k_l,1)})} t'_{h} \mathrm{Tr}\left[\mathrm{Tr}_{\bar{h}}\left[\rho_0\right]\mathrm{Tr}_{\bar{h}}\left[\rho_0\right]\right] -\sum_{\tau=1}^{(L+1)/2} \frac{c'_\tau}{2^{m(2\tau-1)/2}}, \right)^2, \\
\end{split}
\end{equation}
where $c'_{\tau},t'_h \in \mathbb{R}^{+} $ and $P'(S_{(k_u,1)}:S_{(k_l,1)})=\left\{S_{(p,1)} \cup  S_{(p\pm1,1)}\cup S_i| S_i \in P(S_{(k_u,1)}:S_{(k_l,1)}) \right\}$.
If $L$ is even, the variance is written as
\begin{equation}
\label{eq:var_L_layer_edge_even}
\begin{split}
    {\rm{Var}}[k_{PQ}^{(\kappa)}] &=
    \frac{2^{mL}\left(2^{2n_{\kappa}}-1\right)}{(2^{2m}-1)^{2}(2^m+1)^{2(L-1)} 2^{2n_{\kappa}}}\left(2^{m} \sum_{h\in P(S_{(k_u,1)}:S_{(k_l,1)})} t''_{h} \mathrm{Tr}\left[\mathrm{Tr}_{\bar{h}}\left[\rho_0\right]\mathrm{Tr}_{\bar{h}}\left[\rho_0\right]\right] -\sum_{\tau=0}^{L/2} \frac{c''_\tau}{2^{m\tau}}, \right)^2 \\
\end{split}
\end{equation}
with $c''_{\tau},t''_h \in \mathbb{R}^{+} $.
We note that $W_{k_u,1}$ and $W_{k_l,1}$ denote the unitary blocks located at the edge of the light-cone in the first layer.
Also Eqs.~\eqref{eq:var_L_layer} -~\eqref{eq:var_L_layer_edge_even} goes to zero if $\mathrm{Tr}\left[\mathrm{Tr}_{\bar{h}}\left[\rho_0\right]\mathrm{Tr}_{\bar{h}}\left[\rho_0\right]\right]$ is the completely mixed state for any $h$ and reaches the maximum when $\mathrm{Tr}\left[\mathrm{Tr}_{\bar{h}}\left[\rho_0\right]\mathrm{Tr}_{\bar{h}}\left[\rho_0\right]\right]=1$ for all $h$.
We comment that the result for different initial states is also easily obtainable.

Here we summarize key implications obtained from Eqs.~\eqref{eq:var_L_layer} -~\eqref{eq:var_L_layer_edge_even}.
These results indicate that variance of the PQK depends on not only the depth $L$ but also initial state $\rho_0$.
If initial state is a tensor product of arbitrary single-qubit states, i.e., $\rho_0=\sigma_{1}\otimes\sigma_{2}\otimes\ldots\otimes\sigma_{n}$ with an arbitrary single-qubit states $\{\sigma_i\}$, then the variances are $\Omega(2^{-2mL})$ and $\Omega(2^{-mL})$ for case (1) and (2), respectively; this means that it might be possible to preserve the trainability up to $\text{poly}(\text{log}(n))$ depth.
On the other hand, if we prepare an entangled quantum state such as the GHZ state, $\ket{\psi_{GHZ}}=2^{-1/2}(\ket{0}^{\otimes n}+\ket{1}^{\otimes n})$, then the variance would be smaller than the case for a tensor product of arbitrary single-qubit states.
Note that $\mathrm{Tr}[\mathrm{Tr}_{\bar{h}}[\ket{\psi_{GHZ}}\bra{\psi_{GHZ}}]\mathrm{Tr}_{\bar{h}}[\ket{\psi_{GHZ}}\bra{\psi_{GHZ}}]] = 1/2$ for $\bar{h}\neq \emptyset$ or $h\neq \emptyset$ otherwise $\mathrm{Tr}[\mathrm{Tr}_{\bar{h}}[\ket{\psi_{GHZ}}\bra{\psi_{GHZ}}]\mathrm{Tr}_{\bar{h}}[\ket{\psi_{GHZ}}\bra{\psi_{GHZ}}]] = 1$.
In the wort scenario where initial state is random enough to satisfy that $\mathrm{Tr}_{\bar{h}}\left[\rho_0\right]$ is the completely mixed state for almost all $h$, then the variance would be closer to zero. 

\subsection{Variance of the linear projected quantum kernel}\label{ap: variance_sum}

In this subsection, we further check the difference of the variance between PQKs in Eq.~\eqref{eq:def_pqk} and the linear PQK defined as
\begin{equation} \label{eq:def_pqk_linear_re}
\small
    k_{PQ}^{L}(\bm{x},\bm{x'}) = \sum_{\kappa} \mathrm{Tr}\left[ \mathrm{Tr}_{\bar{S}_{\kappa}}\left[\rho(\bm{x},\bm{\theta}) \right] \mathrm{Tr}_{\bar{S}_{\kappa}}\left[\rho(\bm{x'},\bm{\theta}) \right]\right].
\end{equation}
The variance of Eq.~\eqref{eq:def_pqk_linear_re} can be written as
\begin{equation}\label{eq:var_pqkl}
\begin{split}
    {\rm{Var}}[k_{PQ}^{L}] &= {\rm{Var}}\left[\sum_{\kappa} k_{PQ}^{(\kappa)}(\bm{x},\bm{x'}) \right]\\
    &= \sum_{\kappa}  {\rm{Var}}[k_{PQ}^{(\kappa)}] + 2\sum_{\kappa > \kappa'} {\rm{Cov}}[k_{PQ}^{(\kappa)},k_{PQ}^{(\kappa')}], \\
\end{split}
\end{equation}
where ${\rm{Cov}}[A,B]$ represents the covariance of $A$ and $B$.
Then Eq.~\eqref{eq:var_pqkl} means that variance of the linear PQK is different from the simple summation of the variances of Eq.~\eqref{eq:def_pqk_re} because of the covariance terms.
Actually, the covariance of $k_{PQ}^{(\kappa)}$ and $k_{PQ}^{(\kappa')}$ differs depending on whether or not $\kappa$-th qubit(s) and $\kappa'$-th qubit(s) are located in the same subspace of a local unitary block in the last layer.
If $S_\kappa,S_{\kappa'}\subseteq S_{W}$ with a unitary block in the last layer $W$, then we obtain
\begin{equation}
    {\rm{Cov}}[k_{PQ}^{(\kappa)},k_{PQ}^{(\kappa')}] = {\rm{Var}}[k_{PQ}^{(\kappa)}].
\end{equation}
Moreover, if $S_\kappa \subseteq S_{W}$, $S_{\kappa'}\subseteq S_{W'}$ and $S_{W}\cap S_{W'}=\emptyset$, then the covariance reads
\begin{equation}
    {\rm{Cov}}[k_{PQ}^{(\kappa)},k_{PQ}^{(\kappa')}] = 0.
\end{equation}
We notice that the result can be obtained by following exactly what we did for the variance calculation in Appendix~\ref{ap: case_random} and ~\ref{ap: case_ala}; that is the integration of the covariance over the unitary blocks in the last layer.
These results indicate that the variance would not be recovered exponentially by the covariance terms in Eq.~\eqref{eq:var_pqkl}, whereas the variance could possibly increase.

\end{document}